\documentclass{ws-book9x6}
\usepackage{ws-book-thm}
\usepackage{ws-book-har}
\usepackage{caption}
\usepackage{graphicx,subfigure}
\usepackage[margin=0.5in]{geometry}
\usepackage[utf8]{inputenc}
\usepackage{textcomp}
\usepackage{pgfplots}
\usepackage{soul}

\usepackage[pdfpagelabels=false,colorlinks=true,allcolors=black]{hyperref}

\usepackage{bm,mathrsfs,isotope,color,aas_macros}

\makeindex

\begin{document}


\chapter[Title]{Fast Pulsars, Neutron Stars, and Astrophysical Strange Quark Matter Objects}\label{chapter.Farrell}


\author{Delaney Farrell}
  \address{Department of Physics, San Diego State University \\ 5500
    Campanile Drive, San Diego, CA 92182, USA\\ Email:
    dfarrell@sdsu.edu}

\author{Fridolin Weber} \address{Department of Physics, San Diego
  State University \\ 5500 Campanile Drive, San Diego, CA 92182, USA
  \\ Department of Physics, 
  University of
  California at San Diego \\ 9500 Gilman Drive, La Jolla, CA 92093, USA \\ Email:
  fweber@sdsu.edu, fweber@ucsd.edu}

\author{Milva G. Orsaria$^*$, Ignacio F. Ranea-Sandoval$^{**}$, Martín Canullán$^{\S}$}
\address{Grupo de Astrof\'isica de Remanentes Compactos
\\ Facultad de Ciencias Astron\'omicas y
  Geof\'isicas\\ Universidad Nacional de La Plata\\ Paseo del Bosque
  S/N, La Plata (1900), Argentina \\ Consejo Nacional de
  Investigaciones Cient\'ificas y T\'ecnicas (CONICET)\\ Godoy Cruz
  2290, Buenos Aires (1425), Argentina\\ Email:
  $^*$morsaria@fcaglp.unlp.edu.ar, $^{**}$iranea@fcaglp.unlp.edu.ar, $^{\S}$canullanmartin@fcaglp.unlp.edu.ar}

\author{Rodrigo Negreiros} \address{Catholic Institute of Technology \\ 1 Broadway - 14th floor, Cambridge, MA 02142 \\ Instituto de F\'isica,
  Universidade Federal Fluminense, Niter\'oi, Rio de Janeiro,
  Brazil\\ Email: rnegreiros@id.uff.br}

\author{Takashi Katayama$\dagger$, Ian Wagaman$\ddagger$} \address{Department of Physics, San Diego
  State University \\ 5500 Campanile Drive, San Diego, CA 92182, USA
  \\ Email: $^\dagger$tkatayama0396@sdsu.edu, $^\ddagger$iwagaman6967@sdsu.edu}

\clearpage
\vskip 0.5cm

{\bf Abstract:}
This book chapter explores key aspects of neutron stars, pulsar glitches, tidal deformability, fast pulsars, the equation of state, and strange quark matter stars. Challenges in directly measuring neutron star radius have led to reliance on spectroscopic and timing techniques, with uncertainties addressed through careful source selection and theoretical modeling. Pulsar glitches reveal insights into the equation of state through angular momentum transfer within the neutron star. Tidal deformability is crucial in gravitational-wave astronomy, exemplified by the GW170817 event. Fast pulsars, instrumental in astrophysical testing, are classified into ordinary pulsars, millisecond pulsars, and magnetars. The EOS is vital for understanding neutron star internal structure, explored through various models. The chapter delves into the theoretical framework for rotating neutron stars, addressing uniform and differential rotation scenarios and their impacts on mass and radius. Additionally, the intriguing concept of quark stars and strange dwarfs is investigated. The various topics discussed in this book chapter contribute to a broader understanding of dense matter physics, astrophysical phenomena, and the potential for transformative discoveries through advanced observational techniques and technologies like gravitational wave detectors, radio telescopes, and X-ray telescopes.
\vskip 0.5cm

{\bf Keywords:} Neutron stars; Quark stars, Pulsars, Nuclear equation
of state; Strange quark matter, Stellar rotation


\tableofcontents


\renewcommand{\thefootnote}{\arabic{footnote}}

\section{Introduction}\label{sec:Intro}

Neutron stars stand out as the densest observed stellar objects, typically harboring masses within the range of 1 to 2 solar masses (M$_\odot$) and radii spanning from 10 to 15 km. The immense gravitational forces at play within neutron stars compress the matter in their innermost regions to densities exceeding nuclear saturation density ($2.5\times 10^{14}$~g/cm$^3$), turning neutron stars into inherent laboratories for investigating ultra-dense matter \cite{alford01, haensel2007neutron, Alford:2008csi, Becker:2009neutron, Blaschke2018,  Sedrakian:2023PPNP}. Notably, neutron star matter is one of the densest forms of matter in the Universe and also exists in its ground state, characterized by temperatures that are less than $10^{10}$~K (or around 1 MeV), effectively registering as zero temperature on the nuclear scale. The cold, superdense matter encapsulated within neutron stars occupies a unique temperature and density regime that cannot be replicated in terrestrial laboratories, such as the Relativistic Heavy-Ion Collider or the Large Hadron Collider \citep{aad2008atlas}, which are only capable of probing superdense matter in the high-temperature domain. Despite decades of theoretical, experimental, and observational endeavors, understanding the properties of cold, superdense matter remains a forefront area of nuclear and astrophysics.

Determining the equation of state (EOS), which defines the relationship between pressure, density, and temperature of neutron star matter, holds paramount importance
\cite{ozel2016masses,Sedrakian:2007PPNP,Lattimer2016,Burgio2021, Sedrakian:2023PPNP}. This endeavor not only aids in unraveling the nature and characteristics of the strong interaction but also enhances our comprehension of various astrophysical phenomena, including core-collapse supernovae, binary neutron star mergers, and the early Universe.

In addition to the intrinsic significance of neutron stars, a multitude of factors contribute to the evolving landscape of neutron star research. Pulsars, characterized by their intense and precisely timed radiation, offer unique insights into the dynamics of these enigmatic objects. Moreover, the advent of new observational data facilitated by powerful telescopes has significantly advanced our understanding of neutron stars \citep{Kaspi2017}. Notably, the (1) FAST (Five hundred meter Aperture Spherical Telescope) stands out as a remarkable instrument capable of detecting extremely faint radio signals from space \citep{Nan2016}. This capability has rapidly expanded the number of pulsar observations, providing invaluable insights into the behavior of neutron stars. In addition to FAST, several other state-of-the-art observatories and instruments lead the forefront of neutron star research. Positioned on the International Space Station, (2) NICER (Neutron star Interior Composition Explorer): NICER serves as an X-ray telescope meticulously designed to scrutinize the interior composition of neutron stars, delving into the intricacies of their dense cores \citep{nicer_mission}. (3) eROSITA (Extended Roentgen Survey with an Imaging Telescope Array): Functioning as an X-ray instrument on the Spektrum-Roentgen-Gamma mission, eROSITA conducts a comprehensive survey of the X-ray sky \citep{erosita_mission}. Its capabilities significantly contribute to the exploration of neutron stars and other high-energy astrophysical phenomena. (4) Chandra X-ray Observatory: This space telescope, specifically engineered for observing X-rays emanating from high-energy regions of the universe proves invaluable in the study of neutron stars \citep{chandra_mission}. (5) XMM-Newton: Serving as an X-ray observatory designed for high-resolution X-ray spectroscopy, XMM-Newton has played a pivotal role in advancing our understanding of neutron stars and their surrounding environments \cite{xmm_newton_mission}.  (6) VERITAS (Very Energetic Radiation Imaging Telescope Array System): Contributing to the study of neutron stars emitting high-energy gamma rays, VERITAS enhances our exploration of these celestial objects \citep{veritas_collaboration}. (7) LOFAR (Low-Frequency Array): A vast radio telescope array spanning Europe, LOFAR offers unique capabilities for low-frequency radio observations, providing a valuable avenue for studying pulsars and other radio-emitting neutron star phenomena \citep{lofar_collaboration}.  (8) Green Bank Observatory: Instrumental in the examination of pulsars and other radio-emitting neutron star phenomena, the Green Bank Observatory plays a significant role in advancing our understanding of these cosmic entities \citep{GBT_paper}. (9) Jodrell Bank Observatory: The observatory's radio telescopes, including the Lovell Telescope, have been instrumental in observing and monitoring pulsars, aiding in the understanding of these highly magnetized, rotating neutron stars. The precise timing of pulsar signals is crucial for research, including the search for gravitational waves \citep{jodrell_bank}. Last but certainly not least, (9) the Parkes telescope is a state-of-the-art facility for radio astronomy, distinguishing itself in the study of neutron stars and pulsars \citep{Parkes_Telescope}. Collectively, these observatories and instruments represent a comprehensive and cutting-edge approach to unraveling the mysteries of neutron stars.

Beyond conventional neutron stars, the intriguing possibility of strange quark stars, composed of strange quark matter, adds an extra layer of complexity to our exploration.  Strange quark matter is a speculative and exotic form of matter theorized to exist at exceedingly high densities. According to the hypothesis surrounding strange quark matter, it is envisioned as a composition of up, down, and strange quarks -- the fundamental particles constituting protons, neutrons, and hyperons. The proposition suggests that strange quark matter may possess greater stability than conventional nuclear matter, with quarks moving freely rather than being confined within protons and neutrons. Should such matter indeed exist, it could give rise to new categories of compact stellar objects, ranging from compact strange stars, with bulk properties akin to massive neutron stars, to strange quark matter dwarfs, representing the strange counterparts of ordinary white dwarfs. Despite the elusive nature of experimental verification for the existence of strange quark matter, its theoretical implications are pivotal in advancing our understanding of astrophysical phenomena and the fundamental building blocks of matter under ultra-high-density conditions.

In this chapter, we delve into the intricate theoretical modeling of neutron stars, compact strange stars, and strange dwarfs, placing particular emphasis on the influence of rapid stellar rotation. Our investigation encompasses the examination of both differential and rigid body rotation. The primary objective is to comprehensively explore the properties inherent in these stellar objects and discern the pivotal role played by strange quark matter in interpreting the nature of compact stellar entities.

\section{Observation}\label{sec:obs}

Observation of neutron stars has been perhaps the most important catalyst to furthering our knowledge on the subject since their theorized existence. In the last two decades, there has been an increasing amount of high-quality, new data from X-ray and gamma-ray telescopes, including precisely timed observations of pulsars \citep{ozel2016masses}. Today, over 3,000 pulsars have been discovered, and their discoveries have provided valuable constraints on various neutron star properties like mass and radius, cooling \citep{slane2002new}, and magnetic fields \citep{makishima1999cyclotron}. Additionally, these measurements provide constraints on the cold EOS of superdense matter \citep{lattimer2007neutron, dietrich2020multimessenger}, a long-standing mystery in nuclear astrophysics. As we move into an era of new-generation X-ray telescopes, these constraints on neutron star properties and EOS are expected to tighten even further; these discoveries will also aid in our understanding of other astrophysical phenomena like core-collapse supernova, r-process nucleosynthesis, and gravitational wave emission \citep{ozel2016masses}.

\subsection{Observed Neutron Star Properties}

\subsubsection{Mass}

The mass of neutron stars is perhaps the most precisely measured neutron star property, especially those coming from radio pulsar timing of neutron star masses in compact binaries  \citep{RevModPhys.88.021001}. Observed values of masses of neutron stars \citep{neutron_star_masses:Lattimer} are extremely valuable measurements as they test theories of nuclear matter and therefore provide constraints on the nuclear EOS. A notable example of such a constraint resulted from the observation of massive neutron stars. For several decades the canonical mass of a neutron star was 1.4 M$_\odot$, but recent observations of heavier neutron stars ($\geq 2$ M$_\odot$) have widened the range of masses that must be accounted for with theoretical EOS models. Heavy neutron stars directly challenge theoretical models of dense nuclear matter in quantum chromodynamics (QCD). Observations of massive stars enforced the requirement that EOS models for neutron star matter must produce maximum stable masses of at least 2 M$_\odot$. Such models are referred to as "stiff" meaning they have a large pressure for a given density, so observations of massive neutron stars ruled out many softer theoretical EOS models. An example of such massive observations is MSP J0740+6620 with a measured mass of $2.14^{+0.10}_{-0.09}$ M$_\odot$, making it one of the heaviest millisecond pulsars observed thus far \citep{cromartie2020relativistic}. 

Radio pulsars, or highly magnetized, rotating neutron stars, that exist in binary stellar systems generate radio pulses that can be precisely tracked as a method of determining the binary's mass \citep{ozel2016masses}. Radio pulsars have long-term rotational stability, allowing for very precise timing measurements of orbital motion. The binary mass function, $f(M_1, M_2, i)$, can be derived from measurements of five Keplerian parameters: the binary period $P_b$, the projection of the pulsar's orbital semimajor axis $a$ on the line of sight $x \equiv a_i \sin i$, the orbital eccentricity $e$, and time $T_0$ and longitude $\omega_0$ of periastron $\omega$ \citep{Weber:1999book}. The formula for the mass function is given as
\begin{equation}
    f(M_1, M_2, i) \equiv \frac{(M_2 \sin i)^3}{(M_1 + M_2)^2} = \frac{P_b v_1^3}{2 \pi G} = \frac{4 \pi^2 x^3}{G P_b^2},
\end{equation}
where $G$ is Newton's gravitational constant and $v_1 = (2 \pi / P_b)x$ is the orbital velocity of star $M_1$ along the line of sight. Other relativistic effects, like the transverse Doppler effect and gravitational redshift, can also be used to determine the pulsar and companion's masses in a binary system \citep{ozel2016masses, Weber:1999book}. 

Alongside other parameters, mass can also be inferred from stellar emission using other observational techniques like waveform modeling, where a Bayesian inference approach is employed to analyze X-ray waveform oscillations emitted by pulsars \citep{miller2015determining, miller2019psr}. X-ray oscillations can be observed from accretion-powered pulsars, thermal emission of rotation-powered (or non-accreting) pulsars, or some thermonuclear bursts on accreting neutron stars \citep{RevModPhys.88.021001}. It is believed that the X-ray oscillations are emitted from a ``hot spot" on the star's surface, or a region of the stellar surface that is both hotter than its surroundings and offset from the rotational pole of the star. The pulse profile, or waveform, from the hot spot combined with general relativistic effects like light bending and Doppler boosting, can be used to determine the mass and radius of the source. Pulse waveform modeling was pioneered using data from X-ray telescopes like ROSAT and XMM-Newton, but the methodology has become more popular recently with observations from NICER, which has extraordinary absolute timing accuracy when determining the pulse profile of isolated neutron stars and those in binary systems \citep{miller2019psr}. For example, NICER data was used to determine estimations for the mass and radius of the isolated millisecond pulsars PSR J0030+0451 \citep{riley2019nicer,miller2019psr} and PSR J0740+6620 (together with XMM-Newton data) \citep{riley2021nicer, miller2021radius}. The latter study showed that although having a mass $\sim 40 \%$ larger, the radii of PSR J0740+6620 and PSR J0030+0451 are of the same order \citep{riley2019nicer, miller2021radius}. This finding is all the more interesting when considering that the determination of neutron star radius has historically been more difficult than mass, which is discussed further below. 

\subsubsection{Radius}

Historically, direct measurements of neutron star radius do not exist, and inference of radius using various techniques provides large uncertainties when compared to precise mass measurements. Current methods for determining radius commonly rely on the spectroscopic measurements of surface emission to either measure apparent angular size or look for effects of the emission on neutron star spacetime \citep{ozel2016masses, RevModPhys.88.021001}. These methods broadly fall into one of two categories: spectroscopic or timing measurements. 

Spectroscopic measurements from either the surface emission of quiescent low-mass X-ray binary (LMXB) systems or from thermonuclear bursts can be used to constrain the mass-radius relation. If it is assumed that the surface spectrum is a diluted blackbody and the source of the spectrum is at a known distance $D$, the mass-radius (M-R) relation of a slowly spinning neutron star can be defined as 
\begin{equation}
    R^2 \left( 1 - \frac{2 G M}{R c^2}  \right)^{-1} = \frac{F D^2 f_c^4}{\sigma T_c^4},
\end{equation}
where $F$ is the measured flux, $\sigma$ is the Stefan-Boltzmann constant, and $T_c$ and $f_c$ are the color temperature and color-correction factor, respectively \citep{RevModPhys.88.021001}. The color temperature is assumed to be larger than the star's effective temperature, defined as $T_c = f_c T_{\text{eff}}$. The color-correction factor $f_c$ is dependent on both the assumed atmospheric composition and effective surface gravity, where $f_c \approx 1.3 - 2$ \citep{RevModPhys.88.021001}. Under the same assumed conditions, the observed or apparent radius $R_{\text{obs}}$ can be defined as
\begin{equation}
   R_{\text{obs}} =  R \left( 1 - \frac{2 G M}{R c^2}  \right)^{-1/2}.
\end{equation}

The above approximation is reliant on several assumptions and exact measurements and thus may not hold in various instances. In the case of strong magnetic fields, for example, the assumptions of a specific atmospheric composition and a constant surface temperature may be incorrect \citep{ozel2016masses}. For rotating neutron stars that are rotating rapidly (i.e. not spinning slowly), spin-dependent corrections must be applied to the observed angular size as the spacetime can not be described by a Schwarzschild metric due to frame dragging \citep{ozel2016masses}. Additional complications to the approximation are introduced if there is uncertainty when measuring the exact distance $D$ to the star. 

Some uncertainties can be alleviated by choosing sources with low magnetic fields or those that exist in globular clusters of known distances, which is why thermal emission from sources like low-mass X-ray binaries (LMXBs) in quiescence are at the forefront of constraining the neutron star mass-radius relation. The observed thermal spectrum can be fit to a well-motivated theoretical model dependent on the assumed atmosphere of the neutron star in a process called spectral fitting. Many of these theoretical models assume quiescent LMXBs have pure hydrogen atmospheres as the companion stars are hydrogen-rich, but other atmospheric compositions like helium and carbon have also been explored in the literature \citep{steiner2018constraining}. 

As mentioned in the section above, the mass-radius relation can also be explored through X-ray timing techniques like waveform modeling, spin measurements, and asteroseismology. While further explanation of these techniques will not be discussed explicitly, the reader can find more information at \citet{miller2019psr, sotani2011, RevModPhys.88.021001, miller2021radius}.

\subsubsection{Pulsar Glitches and Moment of Inertia}

A pulsar ``glitch", or sudden spin jump, is another observable property that can help constrain our knowledge of the EOS of dense neutron star matter. Rotation-powered pulsars will generally experience a slow and gradual spin down from the emission of magnetic dipole radiation, leading to very regular rotational periods \citep{piekarewicz2014pulsar}. However, some pulsars will experience abrupt spin-ups, leading to a deviation from their otherwise regular rotational periods. 

The mechanism driving these glitches is an angular momentum transfer between the star's solid outer crust and a portion of the liquid interior \citep{link1999pulsar}. A popular model to explain the angular momentum transfer relies on the theory that a pulsar will form vortex lines in the neutron superfluid formed within the inner crust (at densities of 10$^{11}$ to 10$^{14}$ g/cm$^3$). A rapidly rotating pulsar can have superfluid vortices with extremely high areal densities, and the spin up or spin down of the superfluid is dependent on increasing or decreasing the vortex density \citep{alpar1984rapid, piekarewicz2014pulsar}. The superfluid velocity is thus determined by the spatial arrangement of vortex lines \citep{Weber:1999book}. If the vortex lines were assumed to be pinned to a crystal lattice of neutron-rich nuclei, the angular velocity of the superfluid would be largely fixed and appear to be rotating as a rigid body \citep{piekarewicz2014pulsar}. The pulsar will slow down by the emission of magnetic dipole radiation, and as the crust's angular velocity slows down from magnetic torque, the difference between the velocity of the crust and the superfluid regime grows. When this differential rotation between the slower crust and faster superfluid vortices grows too large, some vortex lines become unpinned and suddenly transfer angular momentum to the crust - resulting in a glitch \citep{Weber:1999book}. 

To date, 671 glitches from 224 pulsars have been recorded using high-precision pulsar timing (an extensive and well-maintained list can be found in the Jodrell bank glitch catalog \citep{basu2022jodrell}). While most pulsars have glitches infrequently and irregularly, others like the Vela or Crab pulsar have quasi-periodic glitches \citep{piekarewicz2014pulsar}. An extremely unique case is PSR J0537-6910, the fastest spinning young pulsar recorded, which exhibits large glitches roughly every 100 days \citep{abbott2021constraints}.  

The magnitude of a pulsar glitch is dependent on the star's moment of inertia $I$, another global neutron star property. A pulsar glitch of magnitude $\Delta \Omega_i$ requires the angular momentum
\begin{equation}
    \Delta J_i = I_c \Delta \Omega_i
\end{equation}
where $I_c$ is the moment of inertia of both the star's crust and coupled interior core, estimated to make up 90\% or more of the star's total moment of inertia \citep{link1999pulsar}. Between each glitch, the rotating interior acts as an angular momentum reservoir at a rate of $\dot{J}_{\text{res}}$. This rate is constrained by the average spin-down rate of the crust $\dot{\Omega}$ and the moment of inertia of the reservoir $I_{\text{res}}$ as \citep{link1999pulsar}
\begin{equation}
    \dot{J}_{\text{res}} \leq I_{\text{res}} |\dot{\Omega}|. 
\end{equation} 
The ratio of the moment of inertia of the reservoir $I_{\text{res}}$ and the moment of inertia of the crust $I_c$ provides the lower limit of
\begin{equation}
    \frac{I_{\text{res}}}{I_c} \geq \frac{\Omega}{|\dot{\Omega}|} A = G
\end{equation}
where $A$ is the pulsar's glitch activity parameter and $G$ is the coupling parameter that describes the minimum fraction of $I$ that stores and imparts angular momentum to the crust in the glitch \citep{link1999pulsar, piekarewicz2014pulsar}. The moment of inertia can be expressed in terms of mass and radius, so constraining the upper and lower limits can therefore also provide valuable constraints on the dense matter EOS.  

\subsubsection{Tidal Deformability and Gravitational Waves}

The tidal deformability of compact objects is an important physical quantity for gravitational-wave astronomy. The main reason is that it determines the pre-merger GW signal in binary neutron star merger events. Working up to linear order, the tidal deformability $\lambda$ is given by

\begin{equation}
    \lambda = - \frac{\varepsilon _{ij}}{Q_{ij}},
\end{equation}
\noindent
where $\varepsilon_{ij}$ denotes the external gravitation field produced by the external source and $Q_{ij}$ is the induced mass-quadrupolar moment of the given object. In addition, $\lambda$ is related to the second Love number, $k_2$, and the radius of the compact object, $R$, by

\begin{equation}
    \lambda = \frac{2}{3}k_2R^5.
\end{equation}
\noindent
Finally, the dimensionless tidal deformability, $\Lambda$, is defined by

\begin{equation}
    \Lambda = \frac{\lambda}{M^5},
\end{equation}
\noindent
where the $M$ denotes the gravitational mass of the compact star.
For details on how theoretical calculations related to this quantity are performed, the interested reader can refer to Refs.\ \citet{Hinderer_2008,PhysRevD.99.083014} and references therein.

The detection by LIGO and Virgo observatories of the gravitational waves emitted during event GW170817, a binary neutron star merger (see, for example, \citep{GW170817-new} and references therein) together with its electromagnetic counterpart (see, for example, \cite{2017ApJ...848L..12A}, and references therein) allow to estimate the possible masses, radius and dimensionless tidal deformability of the merging objects. Additionally, the dimensionless tidal deformability of a canonical neutron star was estimated to be $\Lambda_{1.4} =190^{+390}_{-120}$ at a 90\% level with no strong restrictions to the EOS are imposed \citep{GW170817-new}.

\subsection{Fast Pulsars}

Pulsars were first discovered in 1967 by Jocelyn Bell Burnell and Antony Hewish, who detected radio-wave pulses with rapid regularity from an unknown source \citep{hewish1979observation}. Just one year later, Thomas Gold proposed that pulsars are rapidly rotating, highly magnetized neutron stars that emit radio waves from their magnetic poles. Since their discovery, pulsars have served as valuable resources for testing theories of relativistic astrophysics and as a tool for the discovery of astrophysical phenomena like the existence of dark matter \citep{de2021nanograv} or the gravitational wave background \citep{agazie2023nanograv}. To date, there have been over 3,000 pulsars observed, and this number will continue to grow as new radio, X-ray, and gamma-ray telescopes are constructed.

Pulsars can broadly be classified into three categories: ordinary pulsars, millisecond pulsars (MSPs), and magnetars. Ordinary or normal pulsars have rotational periods $P$ between 0.1 to 3 s and magnetic fields of roughly 10$^{12}$ G. This class of pulsars will spin rapidly early in their lifetime ($\lesssim 10^4$ years) but slow as they age, eventually disappearing from view \citep{melrose2017coherent}. MSPs, also referred to as recycled pulsars, are old neutron stars that have been recycled (spun up) by the accretion of mass and angular momentum from a low-mass companion \citep{manchester2017millisecond}. MSPs have spin periods on the order of 1.4 to 20 ms and smaller magnetic fields (B $<$ 10$^{11}$ G) when compared to the other two classes of pulsars. The last class, magnetars, are slowly rotating pulsars with extremely high magnetic fields (10$^{14}$ - 10$^{15}$ G). The radio emission from all three types of pulsars generally tends to fall into a relatively narrow frequency range of $\sim 100$~MHz to 10~GHz, with a smaller subset of pulsars producing high-energy emission. The three categories of pulsars can be differentiated with two measured parameters: the pulse period $P$ and its rate of change $\dot{P}$, which can be used to determine other important quantities like the star's rotational frequency, spin-down rate, luminosity, characteristic age, and braking index \citep{melrose2017coherent}.

As shown by \citet{Glendenning:1997signal}, a potential signal of quark deconfinement in the core of a pulsar could be the alteration in the timing structure of a pulsar's spin-down.  Should quark deconfinement transpire within their cores, the moment of inertia could decrease significantly as the matter in the core undergoes deconfinement to quark matter. This reduction in the moment of inertia leads to an acceleration in rotation, causing an abrupt change in the pulsar's spin-down rate. This transition is identifiable through the examination of the breaking index which would differ drastically from its canonical value of $\sim 3$. Calculations suggest that the braking index may even approach infinity as quark deconfinement is initiated in the stellar core.

As mentioned previously, measuring the stable rotation of pulsars provides valuable constraints on observable properties. The most precise timing measurements come from MSPs, which commonly exist in binary systems with low-density companions like white dwarfs. The small number of isolated MSPs are believed to have lost their companion in some way. For low-mass binaries, this could be from disruptions resulting from the pulsar's strong relativistic winds; for higher-mass double neutron star systems, the orbit could be disrupted by the supernova explosion of the secondary star \citep{lorimer2008binary}.  

The fastest pulsar observed thus far is PSR J1748–2446ad, discovered first in 2004 and confirmed in 2005 \citep{hessels2006radio}. This radio pulsar is located in the globular cluster Terzan 5, which has the largest known population of globular cluster MSPs. PSR J1748–2446ad spins at an impressive frequency of 716 Hz, with a pulse period $P$ of 1.39595482~ms, and a pulse derivative $\dot{P}$ of $\leq 6 \times 10^{-19}$ \citep{hessels2006radio}. Before the discovery of PSR J1748–2446ad, the fastest MSP detected was the first MSP ever detected, PSR 1937+21, spinning at a frequency of 642 Hz \citep{rawley1987millisecond}. A table of various MSPs with spin periods of $\leq$ 2 ms can be found in Table~\ref{tab:pulsars}.

\begin{table}[!ht]
    \centering
    \caption{Sample millisecond pulsars (MSPs) with rotational periods $P$ of $\lesssim 2$~ms not associated with a globular cluster. Included are the rotational periods $P$, year discovered, and the telescope used in the observation (PKS: Parkes Radio Telescope, GBT: Green Bank Telescope, GMRT: Giant Metre Wave Radio Telescope, MK: MeerKat, FAST: Five Hundred Meter Aperture Spherical Radio Telescope). Table adapted from \citep{astropulsar}.}
    \begin{tabular}{l l l l}
    \toprule 
        PSR Name & P (ms) & Year & Telescope \\ \colrule 
        J0034-0534 & 1.88 & 1994 & PKS  \\ 
        J0955-6150 & 1.99 & 2013 & PKS  \\ 
        J1023+0038 & 1.69 & 2009 & GBT  \\ 
        J1036-4353 & 1.68 & 2021 & MK   \\ 
        J1221-0633 & 1.93 & 2017 & GBT  \\ 
        J1227-4853 & 1.69 & 2014 & GMRT \\ 
        J1301+0833 & 1.84 & 2010 & GBT  \\ 
        J1555-2908 & 1.79 & 2017 & GBT  \\ 
        J1723-2837 & 1.86 & 2012 & PKS  \\ 
        J1747-4036 & 1.65 & 2009 & PKS  \\ 
        J1804-2858 & 1.49 & 2018 & PKS  \\ 
        J1810+1744 & 1.66 & 2009 & GBT  \\ 
        J1831-6503 & 1.85 & 2022 & MK   \\ 
        J1833-3840 & 1.87 & 2015 & PKS  \\ 
        J1843-1113 & 1.85 & 2004 & PKS  \\ 
        J1849+0016 & 1.81 & 2020 & FAST \\ 
        J1859+0313 & 1.61 & 2019 & FAST \\ 
        J1902-5105 & 1.74 & 2009 & PKS  \\ 
        J1924+2027 & 1.95 & 2020 & FAST \\ 
        J2036-02   & 1.91 & 2020 & GBT  \\ 
        J2051+50   & 1.68 & 2020 & GBT  \\ \botrule 
    \end{tabular}
    \label{tab:pulsars}
\end{table}

\section{Equation of State}\label{sec:eos}

The essential physics of matter within a neutron star can be succinctly summarized by the EOS, which describes the pressure-density relation within the star. As discussed earlier in the chapter, the matter within the core of a neutron star exists in a low-temperature, high-density regime that cannot be replicated in laboratories or particle colliders. Instead, the EOS of dense neutron star matter must be described by theoretical models that vary widely due to particle composition and mathematical frameworks used. In this section, we discuss the physics of neutron star matter and explore the theoretical implication of the inclusion of quark matter on the EOS. 

\subsection{Nuclear Matter}\label{sec:eos_nuclear}

The density of matter within a neutron star varies greatly from the crust to the core, ranging from a few grams per cubic centimeter at the surface to around $10^{15}~{\rm g/cm}^3$ at the core. The lower density regime spans from the star's surface to its outer and inner crust regions, while the higher density regime comprises the star's core region. As stated in the Introduction, the density within the innermost part of the core region may experience densities exceeding nuclear saturation density ($2.5\times 10^{14}$~g/cm$^3$), giving rise to more exotic types of matter like meson condensates, deconfined quark matter, and strange matter - all of which alter the matter's EOS. 

The least dense portions of a neutron star are its surface and outer
crust region, where matter is compressed to densities up to $10^{6}~
{\rm g/cm}^3$. The surface region is comprised of ordinary atomic nuclei (such as $^{56}$Fe) and non-relativistic electrons and emits the thermal radiation observed by the telescopes discussed in Section \ref{sec:obs}. In the outer crust, at densities of $7 \times 10^{6} - 4.3 \times 10^{11}$~g/cm$^3$, electrons become relativistic and form a relativistic electron gas. Additionally, the atomic nuclei grow more neutron-rich while forming a solid Coulomb lattice. The final portion of the lower-density regime is the inner crust, existing at densities of $4.3 \times 10^{6} - 2 \times 10^{14}$~g/cm$^3$. In the inner crust, atomic nuclei grow extremely neutron-rich to the point neutrons begin to drip out of the nuclei and populate free states; the neutron-rich heavy metals form clusters in a solid lattice and are then immersed in a gas of relativistic electrons and neutrons
\citep{Pethick:PRL.1993,Pethick:NPA.1995,Pethick:1999.conf,Schaffner:PRC.2006,Chamel:LivingRevies.2008}. 

The high-density region of a neutron star, its core, is composed of a relativistic Fermi liquid of neutrons and protons in the conventional description. The liquid is extremely neutron-rich while maintaining a certain fraction of protons, electrons, and muons to maintain chemical or $\beta$-equilibrium. The condition for $\beta$-equilibrium is given as
\begin{equation}
    \mu_B = \mu_n + q_B(\mu_e - \mu_{\nu_e}),
\end{equation}
where $\mu_B$ is the chemical potential of baryons $B$, neutrons ($\mu_n$), electrons ($\mu_e$), and electron neutrinos ($\mu_{\nu_e}$), while $q_B$ is the baryon electric charge. The dense, neutron-rich matter in the core may reach densities of up to 10 times nuclear saturation density, and as nucleons at such high densities begin to overlap, the matter is expected to transition into non-nucleonic states of matter. These states can take several forms: the onset of quark degrees of freedom and deconfined quark matter \citep{ivanenko1965,Itoh1970PThPh,Collins1975PhRvL,Fritzsch_1973, baym1976,keister76,chap77,chap77b,fechner78}, boson condensates formed by pions or kaons \citep{Migdal1972,Sawyer1973,Brown1976,Haensel1982, Mannarelli2019, Kaplan1988, Nelson1987,Brown1994,Lee1994,Knorren1995,Schaffner1996, Glendenning1999,Malik2021,Thapa:2021kfo}, or the formation of other matter with a high degree of strangeness. Due to the weak interactions in the cold and dense neutron star core, strange states of matter can form hyperons  \citep{Ambartsumyan1960SvA,Ambartsumyan1961AZh,Leung1971ApJ,Pandharipande1971NuPhA,Moszkowski1974PhRvD,Bethe1974NuPhA} or color superconducting phases  \citep{rajagopal01,alford01,rischke2004,Alford:2008csi}. Additionally, the existence of self-bound compact stars consisting of strange quark matter has also been discussed in the literature and will be further explained in the next section \citep{Witten:1984cso,terazawa1979,alcock1986,alcock1988,terazawa1991,madsen1999,Weber:2005PPNP,madsen2007}.

While the matter within the core of a neutron star cannot be replicated in terrestrial laboratories, properties of laboratory neutron-rich nuclei can be used to define the EOS for the crust (up to order $10^{11}$~g/cm$^3$). The crust regions are thin, contributing very little to the bulk properties of the star like its mass and radius. Today, two commonly used EOS models for the crust include Baym-Pethick-Sutherland \citep{bps} and Baym-Bethe-Pethick \citep{bbp} (BPS-BBP), or Harrison-Wheler \citep{harrison1965gravitation} and Negele-Vautherin \citep{negele1973neutron} (HW-NV). 

For matter within the core, various theoretical approaches that can broadly be broken into two categories, phenomenological and ab initio, are employed to describe the dense, asymmetric many-body system. Phenomenological models, or relativistic mean-field (RMF) models, are constructed to reproduce properties of finite nuclei and nuclear matter \citep{tong2018relativistic}. For in-depth discussions concerning constraints on the equation of state of high-density matter extracted from nuclear physics and astrophysics, we refer to \citet{Lattimer:2014EPJA, Lattimer2019, stone2021nuclear}.

Ab initio methods instead start with realistic free-space nucleon-nucleon (NN) interactions applied to the nuclear many-body system \citep{sammarruca2008ab}. NN interactions, based on scattering data from free nucleons and properties of the deuteron, are characterized by a repulsive core at short distances, a strong attraction in an intermediate range, and are dominated by the one-pion exchange at large distances \citep{machleidt1987nuclear, machleidt1989adv}. Ab initio methods of this kind include the 
\begin{figure}[htb]
\centering
   \includegraphics[width=7cm]{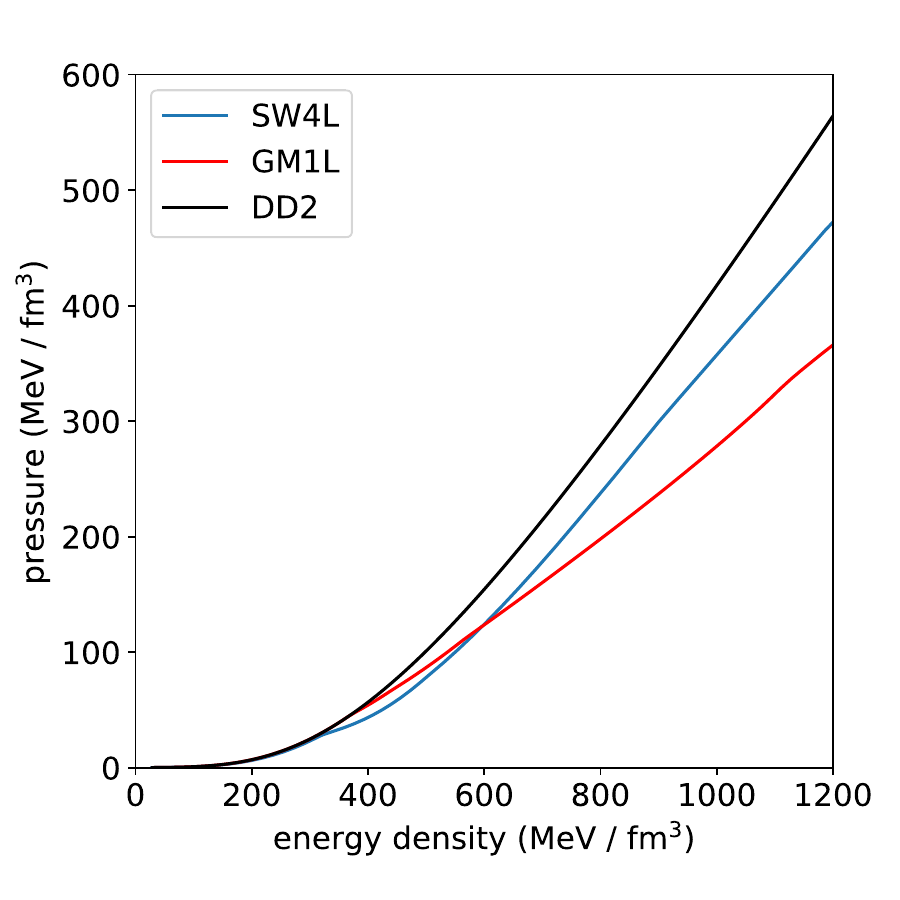}
   \caption{Pressure as a function of energy density of neutron star matter for three nuclear parameter sets: SW4L (containing the $\phi$ and $\sigma*$ mesons), GM1L, and DD2.}
   \label{fig:dd2}
\end{figure}
quantum Monte Carlo method \citep{carlson2015quantum, chen2017abinitio, smith2020quantum}, the self-consistent Green's function method \citep{dickhoff2004self}, and Brueckner-Hartree-Fock (BHF) approaches \citep{terHaar1986equation, terHaar1987nucleons,  poschenrieder1988nuclear, poschenrieder1988PLB, brockmann1990relativistic, sehn1997nucleon, gross1999covariant,van2004relativistic, tong2018relativistic}. For the context of this work, we will focus on the RMF approximation \citep{walecka1974theory, Boguta1977, serot1986recent, reinhard1989relativistic} but a more complete explanation of the varying approximations can be found in \citet{Weber:1999book,glendenning2012compact, Spinella:2020WSBook}. 

In Fig.~\ref{fig:dd2}, we display the nuclear equations of state computed for three prominent RMF parametrization sets: DD2 \citep{Typel:2010, Typel:2018cap}, SW4L \citep{Spinella:2019hns, Malfatti:2020PRC102}, and GM1L \citep{Spinella2017:thesis,Spinella:2018bdq}. The DD2 and GM1L models were recently employed in investigations by \citet{Malfatti:2019PRC100} and \citet{Farrell:WS2023}, where the properties of hot neutron star matter and proto-neutron stars \citep{Pons:1999ApJ} were studied. The Lagrangian is given by (refer also to \citet{Weber:1999book,Glendenning:book2012,Spinella:2020WSBook,Sedrakian:2022astrophysics}):
\begin{eqnarray}
  \mathcal{L} =&& \sum_{B}\bar{\psi}_B \bigl[\gamma_\mu (
    i\partial^\mu - g_{\omega B} \omega^\mu - g_{\rho B}
    {\bm{\tau}} \cdot {\bm{\rho}}^\mu ) - (m_B -
    g_{\sigma B}\sigma) \bigr] \psi_B \nonumber \\
  &&+ \frac{1}{2}
  (\partial_\mu \sigma\partial^\mu \sigma - m_\sigma^2 \sigma^2) -
  \frac{1}{3} \tilde{b}_\sigma m_N (g_{\sigma N} \sigma)^3 -
  \frac{1}{4} \tilde{c}_\sigma (g_{\sigma N} \sigma)^4    \label{eq:Blag} \\
  &&- \frac{1}{4}\omega_{\mu\nu} \omega^{\mu\nu} +
  \frac{1}{2}m_\omega^2\omega_\mu \omega^\mu + \frac{1}{2}m_\rho^2
       {\bm{\rho\,}}_\mu \cdot {\bm{\rho\,}}^\mu -
       \frac{1}{4} {\bm{\rho\,}}_{\mu\nu} \cdot
            {\bm{\rho\,}}^{\mu\nu} \, ,\nonumber
            \label{eq:hadlg}
\end{eqnarray}
where $\psi_B$ represents the various baryon fields, $g_{\sigma B}$, $g_{\omega B}$, and $g_{\rho B}$ are (density-dependent) meson-baryon coupling constants, and $\tilde b_\sigma$ and $\tilde c_\sigma$ denote two additional coupling parameters associated with non-linear (cubic and quartic) self-interactions introduced by \citet{Boguta1977}. The SW4L models incorporate extra $\phi$ and $\sigma^*$ mesons whose Lagrangians are given by \cite{Spinella:2019hns}
\begin{eqnarray}
\mathcal{L}_\phi = -\frac{1}{4}\phi^{\mu\nu}\phi_{\mu\nu} + \frac{1}{2}m_\phi^2\phi_\mu\phi^\mu 
\end{eqnarray}
and
\begin{eqnarray}
\mathcal{L}_{\sigma^*} = \frac{1}{2}(\partial_\mu\sigma^*\partial^\mu\sigma^* - m_{\sigma^*}^2\sigma^*\sigma^*)\, .
\end{eqnarray}
Note that the incorporation of $\phi$ and $\sigma^*$  mesons also modifies the covariant derivative and the baryon $B$ mass terms in the Lagrangian of Eq.\ref{eq:hadlg} by adding $- g_{\phi B} \phi^\mu$ and $- g_{\sigma^* B}\sigma^*$ in the respective parentheses. For all parametrizations, the leptons are described by
\begin{eqnarray}
  \mathcal{L}_l =\bar{\psi}_l\left(i \gamma_\mu \partial^\mu-m_l\right) \psi_l . 
\end{eqnarray}

The saturation properties of symmetric nuclear matter for these models are shown in Table \ref{table:properties}.
\begin{table}[tb]
\tbl{Properties of symmetric nuclear matter at saturation density for
  the SW4L, GM1L, and DD2 nuclear parametrizations.}
{\begin{tabular}{ccccc}\toprule $~~$Saturation property$~~$ &$~~$Units$~~$
    &$~~$SW4L$~~$ &$~~$GM1L$~~$ &$~~$DD2$~~$ \\ \colrule
    $n_0$ &fm$^{-3}$ &0.150  & 0.153 & 0.149\\
    $E_0$ &MeV      &$-16.0$ & $-16.3$ & $-16.02$ \\
    $K_0$ &MeV      &250.0   & 300.0 & 242.7 \\
    ${m^*_N}/{m_N}$  &     &0.70    & 0.70 &0.56 \\
    $J_0$   &MeV       &30.3    & 32.5 & 32.8 \\
    $L_0$ &MeV     &46.5     & 55.0 & 55.3\\
    $U_N$ &MeV     &$-64.6$  &$-65.5$ &$-75.2$ \\ \botrule
    \end{tabular}}
    \label{table:properties}
\end{table}
These properties encompass nuclear saturation density $n_0$, energy per nucleon $E_0$, nuclear compressibility $K_0$, effective nucleon mass $m^*_N/m_N$, asymmetry energy $J_0$, asymmetry energy slope $L_0$, and the value of the nucleon potential $U_N$. The values of $L_0$ listed in Table \ref{table:properties} are consistent with the value of the slope of the symmetry energy deduced from nuclear experiments and astrophysical observations \citep{oertel2017equations}.

The equations of state shown in Fig.~\ref{fig:dd2} include all particles of the spin-$\frac{1}{2}$ baryon octet, comprising the nucleons $(n,p)$ and hyperons 
$(\Lambda,\Sigma^+,\Sigma^0,\Sigma^-,\Xi^0,\Xi^-)$. Additionally, all states of the spin-$\frac{3}{2}$ delta isobar $\Delta(1232)$ ($\Delta^{++},\Delta^+,\Delta^0,\Delta^-$) are taken into account self-consistently. A detailed discussion of the choices for the meson-baryon coupling constants can be found in \citet{Malfatti:2019PRC100,Malfatti:2020PRC102} and in \citet{Spinella:2020WSBook}.

\begin{figure}[htb]
\centering
{\includegraphics[width=9.0cm]{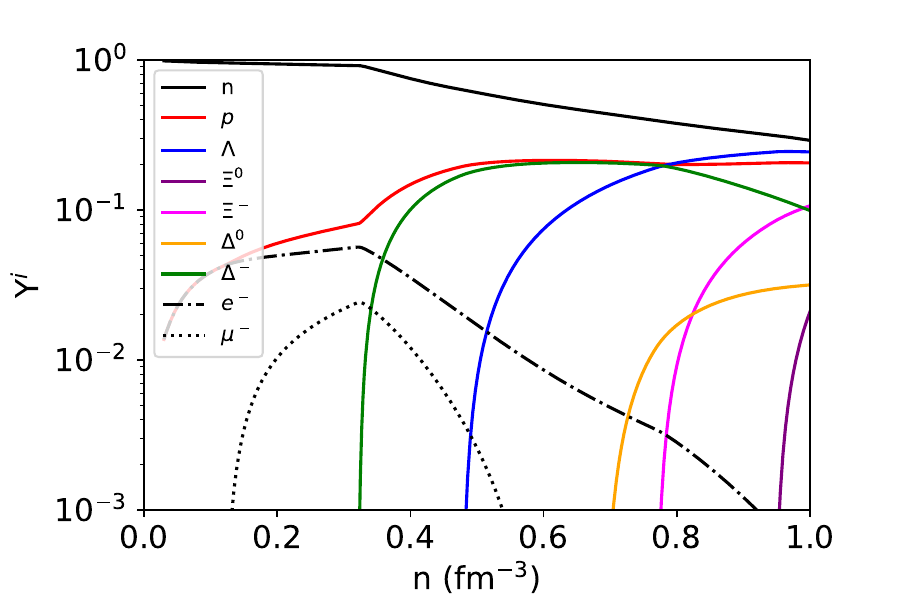}}
\caption{Relative particle fractions $Y^i ~(= n_i/n)$ in cold (T=1~MeV) neutron star matter determined with the SW4L model.}
\label{fig:pop_SW4L_T1}
\end{figure}

\begin{figure}
\centering
{\includegraphics[width=9.0cm]{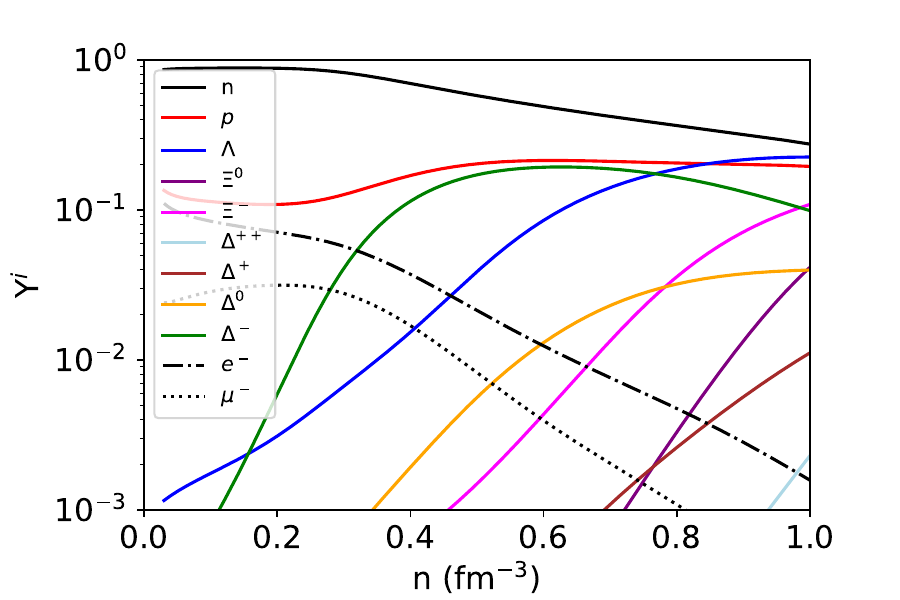}}
\caption{Same as Fig.~\ref{fig:pop_SW4L_T1}, but for neutron star matter at $T=25$~MeV.}
\label{fig:pop_SW4L_T25}
\end{figure}

\begin{figure}[htb]
\centering
{\includegraphics[width=9.0cm]{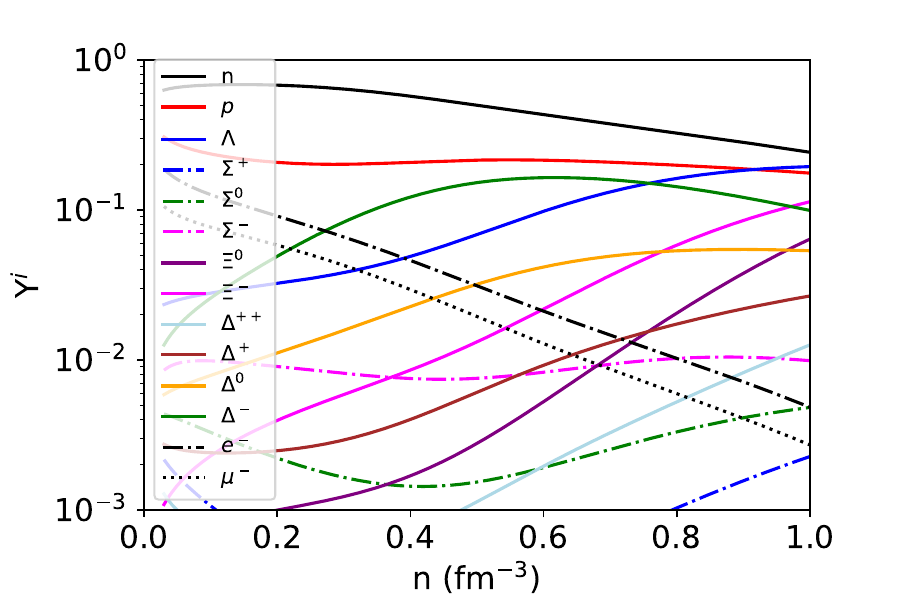}}
\caption{Same as Fig.~\ref{fig:pop_SW4L_T1}, but for neutron star matter at $T=50$~MeV.}
    \label{fig:pop_SW4L_T50}
\end{figure}

\begin{figure}
\centering
{\includegraphics[width=9.0cm]{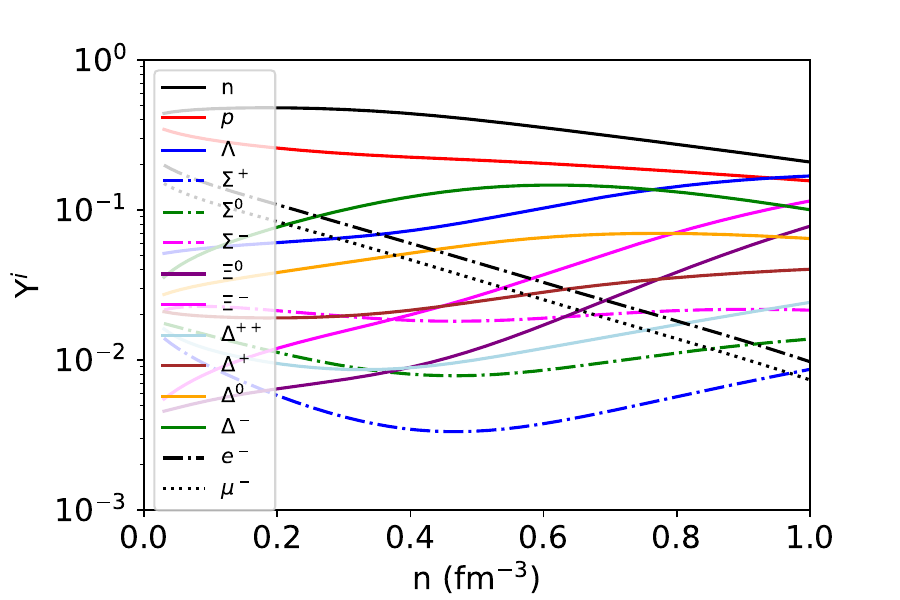}}
    \caption{Same as Fig.~\ref{fig:pop_SW4L_T1}, but for neutron star matter at $T=75$~MeV.}
    \label{fig:pop_SW4L_T75}
\end{figure}

Figures~\ref{fig:pop_SW4L_T1} through \ref{fig:pop_SW4L_T75} showcase the relative particle fractions in dense neutron star matter at temperatures ranging from 1 MeV (representing cold matter) to 75~MeV. These compositions adhere to the constraints of electric charge neutrality and chemical equilibrium, which are maintained
at every density. The delicate equilibrium between positively and negatively charged particles is crucial to ensuring the electric charge neutrality of the matter in the stellar core of a neutron star. Finite temperatures induce alterations in the particle composition within the neutron star core due to increased thermal energy. At higher temperatures, particles gain additional kinetic energy, leading to enhanced thermal motion. This process is governed by fundamental principles such as the Pauli exclusion principle, embodied in the Fermi-Dirac distribution, which dictates the occupation of fermions. The thermal population of states is critically influenced by these principles, impacting the equilibrium conditions that prevail at lower temperatures.

\subsection{Strange Quark Stars and Strange Quark Matter}

Quark stars are hypothetical compact objects in which matter would not be held together by gravitational attraction, but rather by the strong interaction between quarks. Notably, quark stars might possess a lower minimum mass and a smaller radius compared to neutron stars. Additionally, quark stars could feature an exposed quark surface capable of emitting radiation at exceptionally high rates \citep{Usov2001ThermalEF, Zakharov2011} or a thin crust of normal matter \citep{alcock1986, Glendenning:1992crust, Glendenning:1995pnc}.  The existence of quark stars has not been conclusively confirmed, but certain candidates, including RXJ1856.5-37544, PSR J1614-2230, 1E1207, SGR1806, and HESS J1731-347 have been proposed based on observational data \cite{RXu:2003, Turolla:2004bqs,  Weber:2005PPNP, Yuan:2022iua,  Kurban.PLB:2022, Doroshenko:2022asl,horvath2023light}.

The concept of quark stars was initially introduced in the 1970s and 1980s
by several researchers (e.g., \citet{Itoh1970}) and gained prominence among theoretical astrophysicists, driven by the hypothesis that strange quark matter, composed of roughly equal numbers of $u$, $d$, and $s$ quarks, could be more stable than the most stable atomic nucleus, which is $^{62}$Ni. 

In 1971, Bodmer \citep{Bodmer:1971cn} introduced the concept of potentially collapsed atomic nuclei, a crucial element in the exploration of quark matter. The notion of collapsed atomic nuclei and the emergence of new states of matter with greater stability than conventional atomic nuclei were further articulated by Terazawa in 1979 \citep{terazawa1979} and by Witten in 1984 \citep{Witten:1984cso} who proposed that strange quark matter might exhibit greater stability than $^{56}$Fe, the most stable element produced in astrophysical contexts, which is known as strange quark matter hypothesis. We note that it is a standard procedure to compare the energy of strange quark matter to $^{56}$Fe. However, the energy per particle of $^{56}$Fe ranks only third, following $^{62}$Ni and $^{58}$Fe.

Subsequent theoretical studies suggested that the ground state of deconfined quark matter could exist as a color superconductor with quarks forming diquarks similar to Cooper pairs in ordinary matter \citep{rajagopal01, alford01, Alford:2008csi}. 
One conceivable condensation pattern of color superconducting quark matter is the Color-Flavor-Locked (CFL) phase, yet numerous alternative condensation patterns have been proposed in the literature \citep{alford01, Weber:2005PPNP}. If substantiated, this hypothesis could not only establish the reality of quark stars but also indicate their existence as color superconductors.

Building upon this hypothesis, the existence of strange white dwarfs was also proposed \citep{Glendenning:1995pnc}. These would be compact objects distinct from traditional white dwarfs, forming either through the accumulation of normal nuclear matter on their surfaces or by incorporating strangelets found in the galaxy \citep{Glendenning:1995pnc}.

Recently, the special case of the neutron star in the supernova remnant HESS J1731-37 has brought quark stars back into the spotlight \citep{Doroshenko:2022asl}. The estimated mass of this neutron star, M $<$ 0.8 M$_{\odot}$, is much lower than that of any other known neutron star and, in fact, lower than what is believed to be necessary to form a neutron star. 
\begin{figure}[tb]
     \begin{center}
       \subfigure[ Color map for strange quark matter with strong coupling correction.]{%
            \label{fig:window_alpha}
            \includegraphics[width=0.45\textwidth]{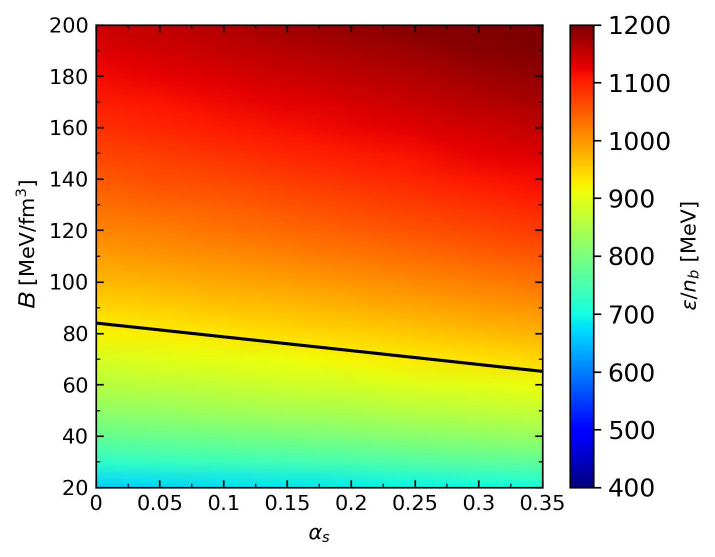}}
          \subfigure[ Color map for CFL quark matter.]{%
           \label{fig:window_delta}
           \includegraphics[width=0.45\textwidth]{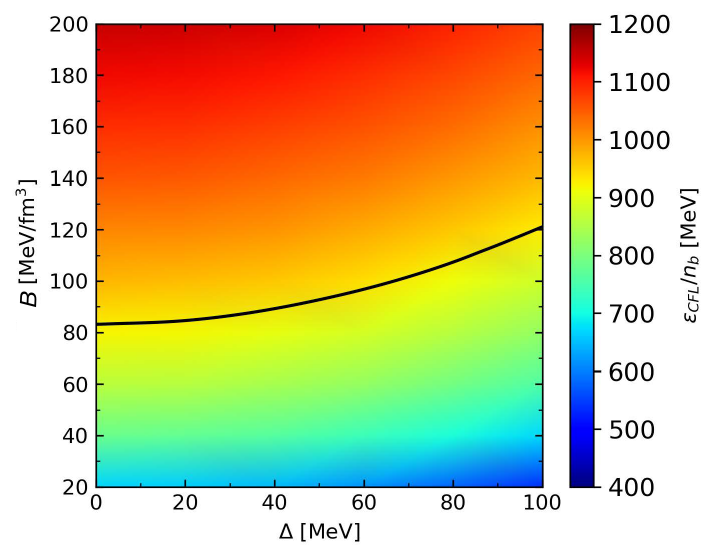}}
    \end{center}
    \caption{Absolute stability of strange quark matter with respect to an $^{56}$Fe nucleus. Energy per baryon is presented as a color map in the $B$-$\alpha_s$ plane, panel (a), ($B$-$\Delta$ plane, panel (b)), for strange (CFL) quark matter within the MIT Bag model (see text for details). The black curve represents the constraint corresponding to the $^{56}$Fe mass. Only $B$-$\alpha_s$ and $B$-$\Delta$ pairs below the black curves result in combinations for EOS that lead to stable astrophysical configurations.}
   \label{fig:stab_win}
\end{figure}
Therefore, the question arises of how such a light object could have formed and whether it is indeed a neutron star or a more exotic object that lost part of its mass in a transition from a neutron star to a quark star.


In this context, we dedicate part of this chapter to these exotic and hypothetical compact objects. Numerous studies investigating the properties of quark stars have been conducted over the past several decades, with a renewed interest in the topic following observations like HESS J1731-37 and the detection of gravitational waves from stellar binary mergers. Recently, the first fully general relativistic modeling of a merger of two quark stars was conducted \cite{zhu2021fully}, which discusses the dynamical mass lost \cite{zhu2021fully} and the threshold mass for prompt collapse into a black hole for a quark star binary \cite{zhou2022evolution}. For a recent examination of the stability of rotating quark stars, we direct readers to the works of \cite{Zhou2021, chen2023fully} and the references contained therein.  

\subsubsection{Equation of State of Quark Matter and Stability Windows}\label{quark_eos_sec}

To construct self-bound quark stars, we will use two descriptions for the EOS of the matter that composes them: quark matter with corrections due to the strong coupling constant, and quark matter in color superconducting CFL state.

Corrections in bulk quark matter due to strong coupling constant, $\alpha_{s}$,  were included by \cite{Farhi:1984sm} in 1984. In the case where the quark gas interacts strongly, $O(\alpha_s)$ corrections (being $\alpha_s$ the strong coupling constant among quarks) must be made through the perturbative expansion of QCD, developing an improved renormalization group. To do this, a renormalization point $\rho$ where $m(\rho)$ and $\alpha_s (\rho)$ are defined must be chosen. Specifically, $\rho=M_N/3=313$~MeV was chosen, where $M_N$ is the mass of the neutron. Thus, the pressures for light quarks are as follows:
\begin{eqnarray}
    P_u &=& \frac{{\mu^4_u}}{4{\pi}^{2}} (1-\frac{2\alpha_s}{\pi}) \, , \\
    P_d &=& \frac{{\mu^4_d}}{4{\pi}^{2}} (1-\frac{2\alpha_s}{\pi}) \, ,
\end{eqnarray}
\begin{align}
    P_s=\frac{1}{4{\pi}^{2}}\biggl\{\mu_s {(\mu_s^{2}-m_{s}^{2})}^{1/2}(\mu_s^{2}-\frac{5}{2}m_{s}^{2})+\frac{3}{2}m_{s}^{4}\, \ln{\frac{\mu_s + {(\mu_s^{2}-m_{s}^{2})}^{1/2}}{m_{s}}} \nonumber \\
   -  \frac{2\alpha_s}{\pi}\biggl[3{\biggl(\mu_s {(\mu_s^{2}-m_{s}^{2})}^{1/2}-m_{s}^{2}\, \ln{\frac{\mu_s + {(\mu_s^{2}-m_{s}^{2})}^{1/2}}{\mu_s}}\biggr)}^{2} \nonumber \\
  -   2{(\mu_s^{2}-m_{s}^{2})^{2}}+3m_{s}^{4}{\ln{(\frac{m_{s}}{\mu_s}})}^{2}
   +6\, \ln{(\frac{\rho}{\mu_s})}\biggl(\mu_s m_{s}^{2} {(\mu_s^{2}-m_{s}^{2})}^{1/2} \nonumber \\
  -  m_{s}^{4}\, \ln{\frac{\mu_s + {(\mu_s^{2}-m_{s}^{2})}^{1/2}}{m_{s}}}\biggr)\biggr]\biggr\} \, ,
\end{align}
where we have considered massless $u$ and $d$ quarks, the strange quark mass, m$_s$ = 96 MeV; $\mu_i=\mu+q_i \mu_e$ is the chemical potential of the quark $i$, with electric charge $q_i$ in $\beta-$equilibrium and $\mu_e$ is the chemical potential of the electron. The contribution of the electrons to the pressure is given by
\begin{equation}
   P_e = \frac{{\mu^4_e}}{12{\pi}^{2}} \,.
\end{equation}
For simplicity, muons are not included because their contribution to quark matter at zero temperature is negligible. The equation of state for quarks with corrections due to the strong coupling constant, within the MIT Bag model, arises from a mechanical-statistical or thermodynamic analysis in which the energy density is given by
\begin{align}
    \varepsilon = \sum_{i=u,d,s,e} \left( {\mu}_i\,  n_i - P_i \right) + B \, ,
    \label{EoSq}
\end{align}
where \textit{i} ranges over the three quark flavors, and electron, arising from the condition of chemical equilibrium among the system's particles. The bag constant, $B$, acts as a vacuum energy per unit volume (or pressure) and takes into account the long-range interactions of QCD. The pressure due to the contribution of the three quarks in the CFL phase is given by \citep{smith:2009mic}
\begin{equation}
P_{\rm CFL} = \frac{3\mu^4}{4\pi^2} \, (1-c) - \frac{3\mu^2}{4\pi^2} \, 
(m^2_s - 4\Delta^2)-B \, ,
\label{eq:19}
\end{equation}
where $c = 2\alpha_s/\pi$ and $\Delta$ is the color superconducting gap, treated here as a free parameter. We will consider c=0 for the CFL phase. The energy density in this case is given by
\begin{equation}
\varepsilon_{\rm CFL}=-P_{\rm CFL}+\sum_{i=u,d,s} \mu_i n_i,
\end{equation}
where $n_i = \partial P_{\rm CFL}/\partial \mu_i$.

It is worth noting that matter in a CFL phase has properties similar to those of a superfluid \citep{Alford:2008csi}. In our phenomenological treatment of this phase, we have assumed electric and color charge neutrality. We have 
mentioned that quarks pair up to form Cooper pairs, implying that all quarks of different spins and flavors have the same Fermi momentum and, therefore, the same number density. This leads to the CFL phase being electrically neutral on average, allowing us to consider the presence of electrons negligible. While each Cooper pair contains one quark of each of the three fundamental colors, we can also assume that the color combination in each Cooper pair is such that the total color charge of the CFL phase is zero.

As mentioned earlier, the hypothesis of absolute stability of quark matter suggests that this matter is the ground state of hadronic interactions, i.e., it is more stable than nuclear matter. In this definition, the critical pressure is zero: the true ground state of matter is always quark matter. Figure~\ref{fig:stab_win} shows the stability window of quark matter considered in this chapter. The black line in both figures corresponds to the mass of the most stable known element in nature that can be produced in astrophysical contexts, iron $^{56}\rm{Fe}$. Any combination of parameters below the black line in panels (a) and (b) of Fig.~\ref{fig:stab_win} gives us an appropriate equation of state for the construction of self-bound stars.

\begin{figure}[tb]
    \includegraphics[width=13.5cm]{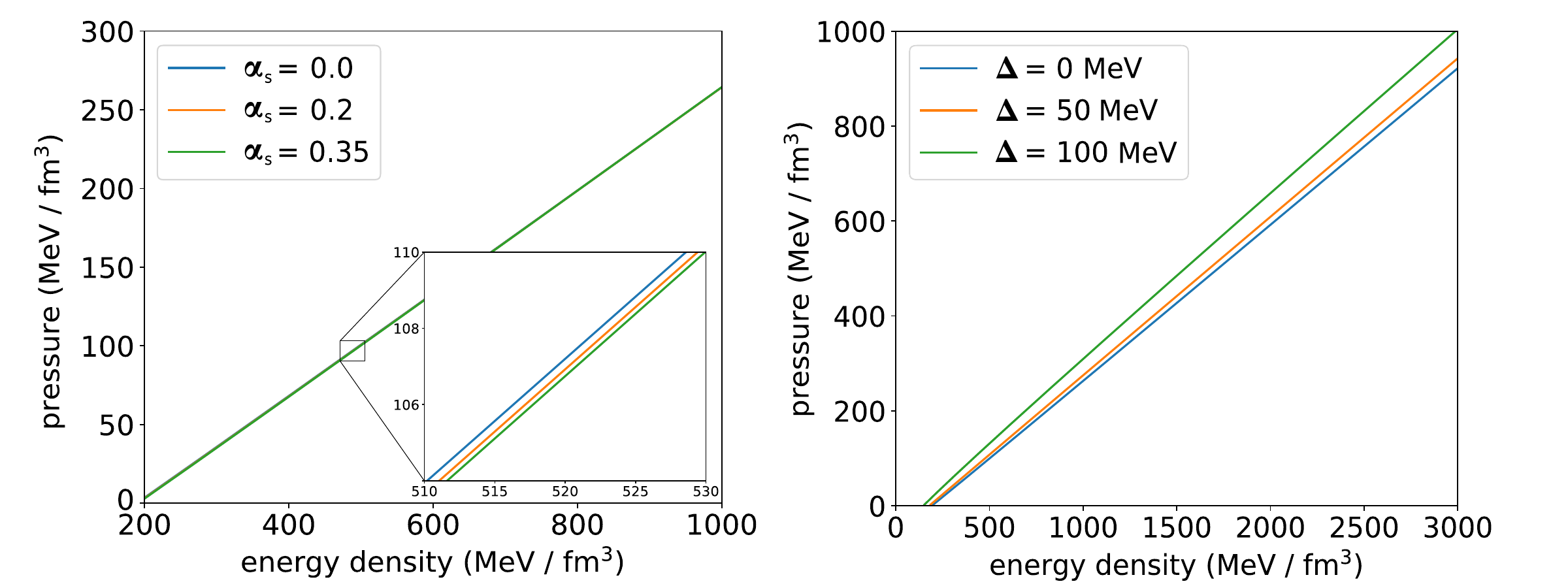}
    \caption{Pressure as a function of energy density for quark matter modeled with the MIT Bag Model, where $B = 45$~MeV/fm$^3$. The right panel demonstrates the impact of modifying the strong coupling constant $\alpha_s$, while the left panel shows variations in the superconducting gap $\Delta$ within the Color-Flavor-Locked (CFL) phase. (See text for details). }
    \label{fig:quarkeos}
\end{figure}

The EOS for the two models of quark matter considered in this chapter are shown in Figure~\ref{fig:quarkeos}. Increasing the strong coupling constant $\alpha_s$ or the superfluid gap $\Delta$ leads to a stiffened quark EOS: for the same energy density, the pressure is higher when these two parameters increase.


\section{Properties of Strange Dwarfs}\label{sec:strange}

\begin{figure}
    \centering
    \includegraphics[width=9cm]{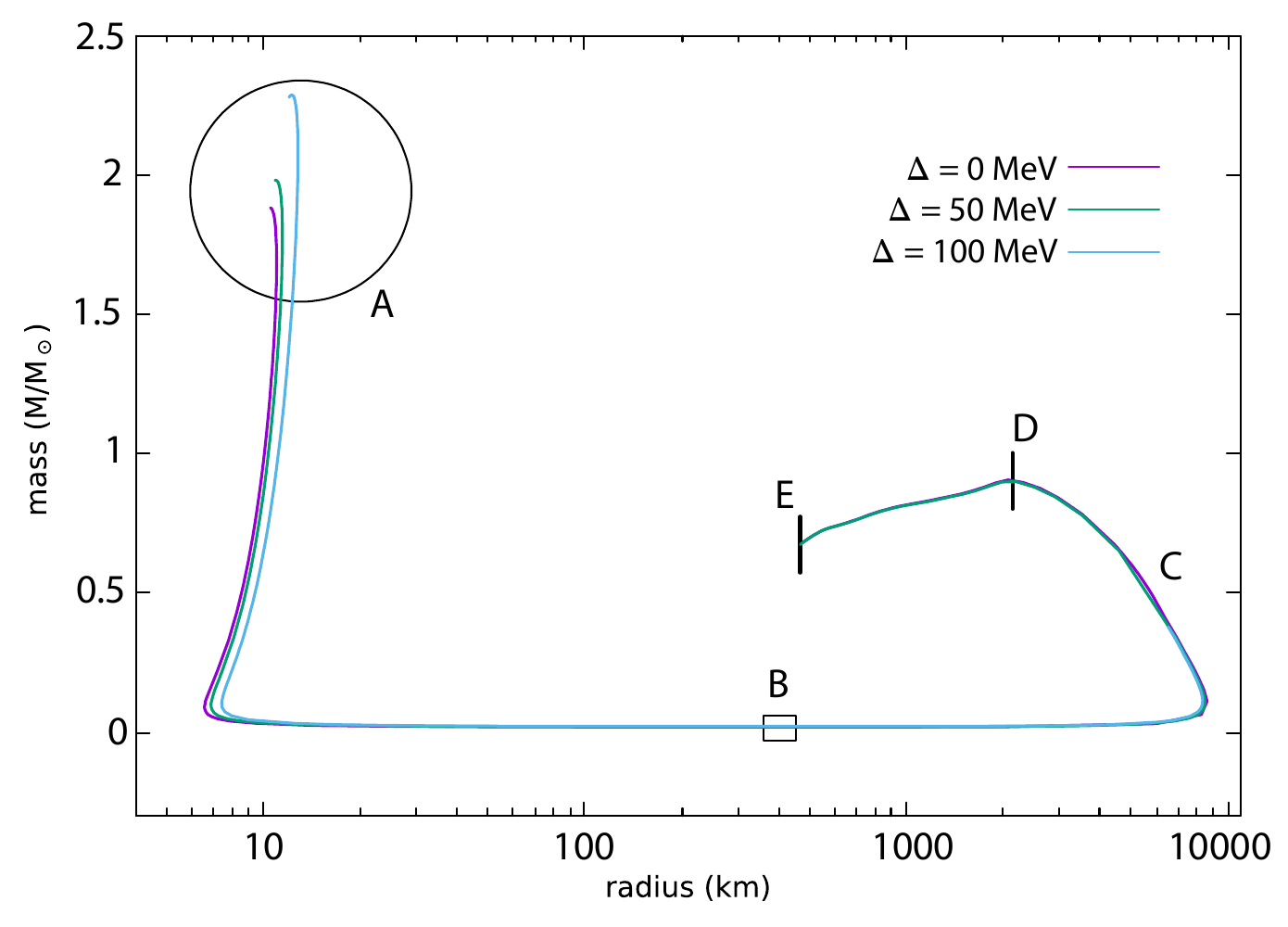}
    \caption{Mass-radius relations for quark stars constructed with the MIT Bag Model varying the CFL superfluid gap $\Delta$. The nuclear crust is described by the Baym-Pethick-Sutherland EOS. 
    (See text for details.) }
    \label{fig:quark_crust}
\end{figure}

Continuing our exploration, we now direct our focus toward the unique characteristics of hypothetical strange dwarfs. These intriguing objects bear a resemblance to white dwarf stars but distinguish themselves by possessing cores composed of strange quark matter, which are enclosed within a 
conventional matter crust. Figure \ref{fig:quark_crust} illustrates the comprehensive mass-radius relationship of strange quark matter stars, encompassing a wide range from the compact members ('A') to the enigmatic strange dwarf objects ('C' to 'E'). The minimum masses of these sequences are confined within the rectangular area labeled 'B.' Notably, at 'D,' we 
find the maximum-mass white dwarfs. However, it's essential to note that stars between 'D' and 'E' are prone to unstable radial oscillations. At point 'E,' the strange matter core at the center of white dwarfs has effectively shrunk to zero, resulting in the transformation into an ordinary white dwarf situated within the oscillatory unstable region.

The stability of strange dwarfs against radial perturbations was called into question by Alford and their collaborators in 2017, casting uncertainty on the potential existence of such celestial bodies \citep{Alford2017}. Subsequent investigations by various research groups \citep{DiClemente2023,  goncalves2023revisiting} have shed light on this matter, demonstrating that strange dwarfs can indeed achieve stability under certain circumstances, provided that the transitions between quarks and hadrons occur at a slower rate than the radial perturbations themselves. Here, we proceed with the presumption that these conditions hold.
\begin{figure}
    \centering
    \includegraphics[width=9cm]{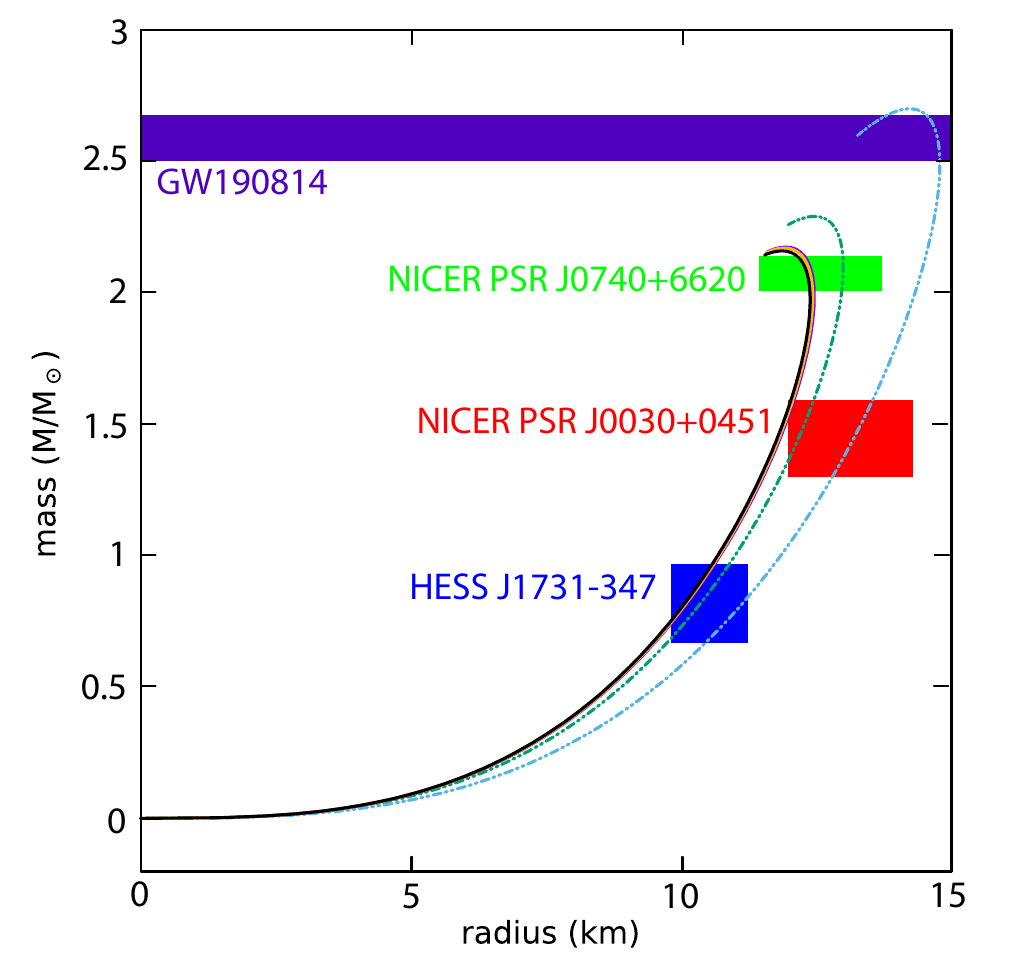}
    \caption{Mass-radius relations for quark stars constructed within the MIT Bag Model, computed for different coupling constants $\alpha_s$ (shown with solid lines) and superfluid gaps $\Delta$ (shown with dashed lines), with $B$ = 45 MeV/fm$^3$, for the quark EOS models considered in this chapter. The results are compared with recent compact-star observations.}
    \label{fig:quark_obs_compare}
\end{figure}

While it is not anticipated that strange matter endured the extreme
conditions prevalent in the early Universe \citep{Alcock.PRD:1985}, there exist alternative scenarios for its generation. One such possibility is the formation of strange matter through the gravitational collapse of stellar cores during supernova explosions, with subsequent dissemination into the interstellar medium occurring during the merger of neutron stars with other compact objects \citep{Bucciantini.PRD:2022}. Despite these potential sources, strange stars and strange dwarfs have, thus far, eluded conclusive detection due to their striking resemblance to neutron stars and white dwarfs, respectively. Recently, reports have emerged suggesting possible candidates for strange dwarfs \citep{Kurban.PLB:2022}. Nevertheless, this situation may witness a transformative shift in the forthcoming years, chiefly owing to the advancements in gravitational wave astronomy. In particular, Perot {\it et al.}  \citep{Perot.PRD:2023}, have demonstrated the feasibility of identifying hidden strange dwarfs within white dwarf binary systems. This can be achieved through precise measurements of their tidal deformability utilizing future space-based gravitational-wave detectors, such as the Laser Interferometer Space Antenna (LISA). These groundbreaking developments hold the promise of unraveling the mysteries surrounding these novel celestial objects \citep{Perot.universe:2023}.

\section{Properties of Rotating Neutron Stars}\label{sec:rotation}

Neutron stars can exhibit rigid body rotation, wherein the entire star
rotates as a singular entity, maintaining a consistent rotation
rate. Alternatively, neutron stars may display differential rotation,
with different layers of the star rotating at varying rates.  As
neutron stars cool over time, diverse mechanisms come into play,
leading to the damping of differential rotation and the establishment
of uniform rotation. A crucial process in this transition is Eckman
pumping, a phenomenon wherein the interaction between the star's
magnetic field and its interior matter creates a torque that aligns
the rotation axis of the star with its magnetic axis. This alignment
process aids in reducing differential rotation by redistributing
angular momentum within the star. The combination of angular momentum
conservation during the collapse phase and subsequent processes like
Eckman pumping guides the evolution of neutron stars from initially
hot and differentially rotating objects to cooler ones with more
uniform rotation. Understanding these processes is essential for
comprehending the internal dynamics, magnetic field evolution, and the
overall lifecycle of neutron stars.

In this section, we provide a brief overview of the theoretical
framework that elucidates the behavior of rotating neutron stars. We
present results detailing the properties of neutron stars, focusing on
both differentially rotating and uniformly rotating scenarios. 

\subsection{Uniform Rotation}

To calculate neutron star properties, the stellar structure equations must be solved in the framework of Einstein's theory of general relativity. These equations are derived from Einstein's field equation and depend on the stellar matter's equation of state, or the underlying relationship between pressure $P$ and energy density $\epsilon$.

The simplest form of the stellar structure equations describes non-rotating, spherically symmetric objects in hydrostatic equilibrium. In this case, the stellar structure simplifies to an ordinary differential equation describing the pressure gradient within the star, known as the Tolman-Oppenheimer-Volkoff (TOV) equation (given in geometrical units where $G = c = 1$):
\begin{equation}
\label{eq:tov}
    \frac{dP}{dr} = -\frac{(\epsilon+P)(m+4\pi r^3 P)}{r^2\left(1-\frac{2m}{r}\right)},
\end{equation}
where $m$ is the gravitational mass enclosed within a sphere of radius $r$, $P$ is the pressure, and $\epsilon$ is the energy density. When integrating the TOV equation, one first would specify a value for the central density, $\epsilon_c$, which in turn is used to determine the pressure value at the center of the star, $P_c$, using the specified EOS; at the center of the star ($r = 0$), the mass $m$ and pressure differential $P'$ disappear. For an infinitesimal increase in $r$, the mass of the star can be solved for as
\begin{equation} \label{eq:2}
    \frac{dm}{dr} = 4\pi r^2 \epsilon.
\end{equation}
The values $\epsilon_c$, $P_c$, and $m$ are used to solve for $P'$, thus allowing for the determination of $P$ at the next radial step, which thus determines $\epsilon$ from the EOS. The computation of these parameters continues successively, increasing $r$ until the pressure vanishes at the edge of the star. Then, the total gravitational mass $M$ of a star can be deduced from the full radius $R$ of the star as $M \equiv m(R) = 4\pi \int_0^R dr ~r^2\epsilon$. 

Rotating neutron star calculations are performed in the framework of general relativity and depend on the matter's nuclear equation of state. Modeling rotating neutron stars is more complicated than spherically symmetric, non-rotating stars for a few reasons: rotation deforms neutron stars, stabilizes them against collapse, and drags along the local inertial frames inside and outside of them so that they co-rotate with the stars \citep{mellinger2017quark, Weber:1999book}. Rapid rotation can cause deformation where the star flattens at the pole and radially expands in the equatorial direction; this deformation results in the stellar structure equations not only being dependent on the radial coordinate $r$ but also on the polar coordinate $\theta$. Rotation is a physical mechanism that stabilizes massive stars against collapse, resulting in rotating neutron stars being able to sustain up to 20\% more mass than their non-rotating counterparts \citep{kalogera1996maximum}. The increased mass of rotating stars alters the geometry of spacetime by introducing a dependence on the star's rotational frequency to the line element and self-consistency condition to the stellar structure equations to account for the dragging of local inertial frames \citep{Weber:1999book}.   

To solve for properties of rotating stars, we begin with the metric, which has the form
\begin{equation}
    ds^2 = -e^{\gamma-\rho} dt^2 +  e^{2\alpha}(dr^2 +r^2d\theta) + e^{\gamma-\rho}r^2 \sin^2\theta (d\phi - \omega dt)^2,
\end{equation}\label{eq:metric}
where $\gamma$, $\rho$, $\alpha$, and $\omega$ are metric functions dependent on the radial coordinate $r$ and polar angle $\theta$, and $\omega$ describes frame dragging due to rotation. These functions also implicitly depend on the star's angular velocity $\Omega$. The metric functions are computed from Einstein's field equation
\begin{equation}
    R^{\kappa \sigma} - \frac{1}{2} Rg^{\kappa \sigma} = 8\pi T^{\kappa \sigma},
    \label{eq:Efe}
\end{equation}
where $R^{\kappa \sigma}$ is the Ricci tensor, $R$ is the curvature scalar, and $g^{\kappa \sigma}$ is the metric tensor. $T^{\kappa \sigma}$ is the energy momentum tensor given by
\begin{equation}
    T^{\kappa \sigma} = (\epsilon + P) \, u^\kappa u^\sigma + g^{\kappa \sigma} P,
\end{equation}
where $\epsilon$ and $P$ are supplied by the equation of state. 
The partial differential equations for the metric functions $\gamma$, $\rho$, $\alpha$, and $\omega$, which follow from Eq.~(\ref{eq:Efe}), are to be solved self-consistently for a given rotational stellar frequency \citep{Paschalidis2017}. The absolute constraint on rapid rotation is defined by mass shedding, occurring at the Kepler frequency, $\Omega_K$. However, this limit might not be reached due to instabilities driven by gravitational radiation-reaction-driven instabilities known as $r$ and $f$-modes \citep{Andersson:2001IJMPD,Paschalidis2017, Bratton2022}. Certain classes of instabilities may be associated with the emission of gravitational waves from neutron stars \cite{Orsaria.2019}. 

The Kepler frequency is determined by the centrifugal force and gravity. In classical Newtonian mechanics, the Kepler frequency can be defined as 
\begin{equation}
    \Omega_K = \sqrt{\frac{M G}{R^3}},
\end{equation}
but this formula is drastically modified in general relativity. Following the metric function shown in Eq.~(\ref{eq:metric}), the general relativistic Kepler frequency is defined in terms of the metric functions as
\begin{equation}
    \Omega_K = \frac{e^\rho}{r} V + \omega
    \label{eq:Omega_K}
\end{equation}
where the quantity $V$ denotes the equatorial velocity given by
\begin{equation}
    V = \frac{\omega'}{(8 + \gamma' - \rho')} r e^{-\rho} + \sqrt{\frac{\gamma' + \rho'}{(8 + \gamma' - \rho')} + \left( \frac{\omega'}{(8 + \gamma' - \rho')} r e^{-\rho} \right)^2}.
    \label{eq:V_eq}
\end{equation}
Equations (\ref{eq:Omega_K}) and (\ref{eq:V_eq}) are to be evaluated self-consistently at the star's equatorial radius. 

Once a set of self-consistent solutions are found for the metric potentials, bulk properties of neutron stars like mass, radius, and $\Omega$ can be determined.

\subsubsection{Results for Uniformly Rotating Compact Objects}\label{subsec:uniform}

\begin{figure}
     \includegraphics[width=13.5cm]{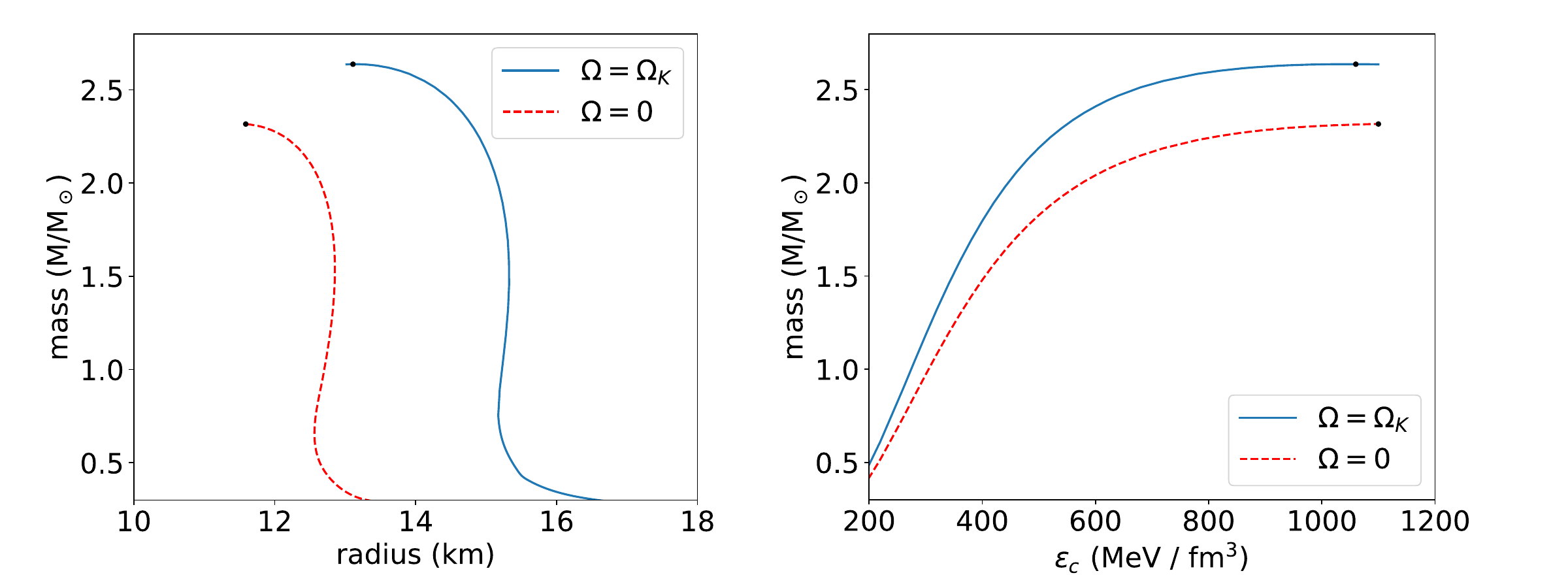}
    \caption{Gravitational mass versus radius (left) and gravitational mass versus central stellar density (right) for both non-rotating ($\Omega=0$) neutron stars and neutron stars rotating at the Kepler frequency ($\Omega=\Omega_K$), modeled with the DD2 nuclear EOS. Mass peaks are highlighted with black dots.}
   \label{fig:prns_dd2}
\end{figure}

This section presents comprehensive insights into the bulk properties of uniformly rotating neutron stars and quark stars. The calculations for uniformly rotating neutron stars utilize the nuclear DD2 model, as detailed in Section~\ref{sec:eos_nuclear}. Figure~\ref{fig:prns_dd2} provides results for both a sequence of non-rotating stars, computed using the TOV equation, as expressed in Eq.~(\ref{eq:tov}), and a sequence of uniformly rotating stars at their mass-shedding limits ($\Omega = \Omega_K$).

In Figure~\ref{fig:prns_dd2}, the left pane illustrates the mass-radius relationships for both non-rotating and uniformly rotating stars. Correspondingly, the right pane of Fig.~\ref{fig:prns_dd2} presents the gravitational mass as a function of central energy density, with each sequence's maximum mass denoted by a solid black dot.
The mass peak values for both the non-rotating and uniformly rotating sequences derived from the DD2 model are 
\begin{table}[]
\centering
\caption{Maximum gravitational mass with corresponding radius and central density of non-rotating and uniformly rotating neutron stars modeled with the DD2 parameter set.}
\begin{tabular}{l c c c}
\toprule
  Rotation & Mass [M$_\odot$] & Radius [km] & $\epsilon_c$ [MeV/fm$^{3}$]  \\ \colrule
  Non-rotating (TOV)   & 2.316 & 11.589 & 1100  \\
  Uniform ($\Omega_K$) & 2.637 & 13.111 & 1060   \\ \botrule
\end{tabular}
\label{tab:rotation_uniform}
\end{table}
provided in Table~\ref{tab:rotation_uniform}, along with their respective radii and central energy density values. Notably, uniform rotation at the Kepler limit enhances the mass peak for the DD2 model by around 14\%. 

\begin{figure}
     \includegraphics[width=13.5cm]{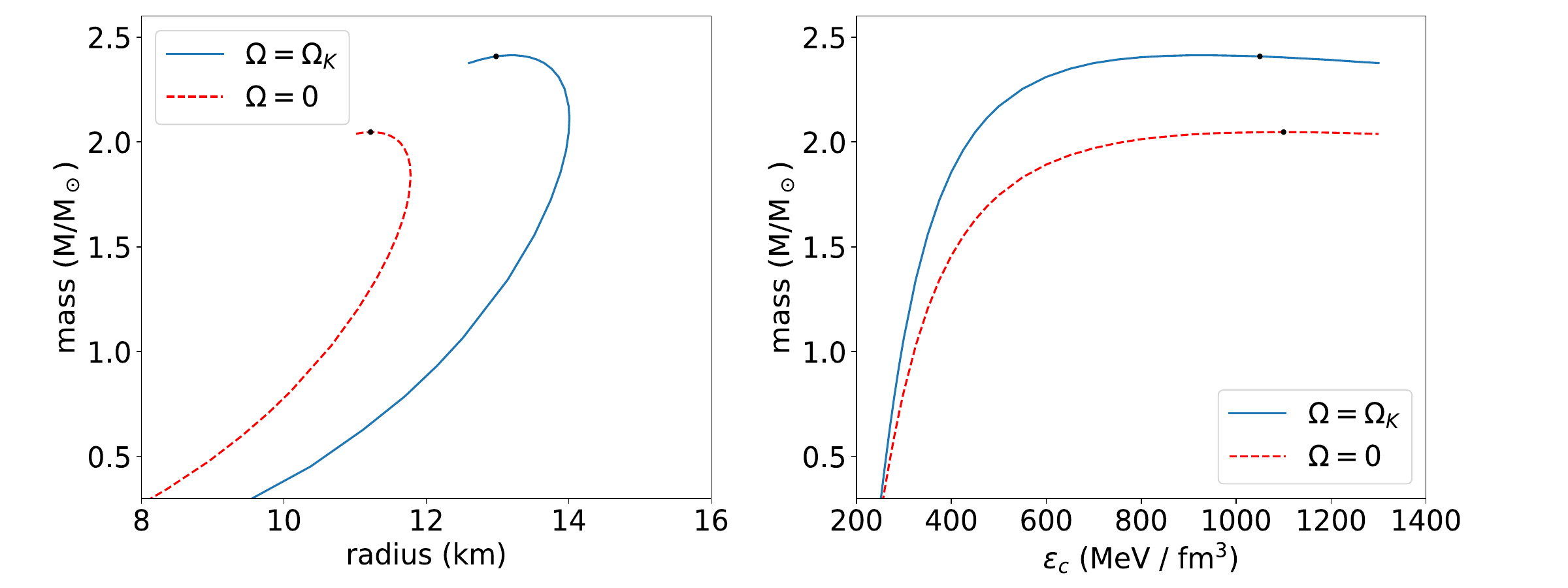}
    \caption{Gravitational mass versus radius (left) and gravitational mass versus central density (right) of non-rotating ($\Omega=0$) quarks stars and quark stars rotating at the Kepler frequency ($\Omega=\Omega_K$), for quark matter modeled within the MIT Bag Model with $B = 45$~MeV/fm$^3$. The mass peaks are indicated with black dots.}
   \label{fig:prns_quark}
\end{figure}

To calculate the properties of uniformly rotating quark stars, the EOS for quark matter is modeled within the MIT Bag Model with a bag constant $B = 45$~MeV/fm$^3$, as explained in Section~\ref{quark_eos_sec}. As for the  DD2 model, a sequence of non-rotating stars and a sequence of uniformly rotating stars at their mass-shedding limit ($\Omega = \Omega_K$) for quark matter are shown in Figure~\ref{fig:prns_quark}. 

\begin{figure}[htb]
\centering
   \includegraphics[width=8cm]{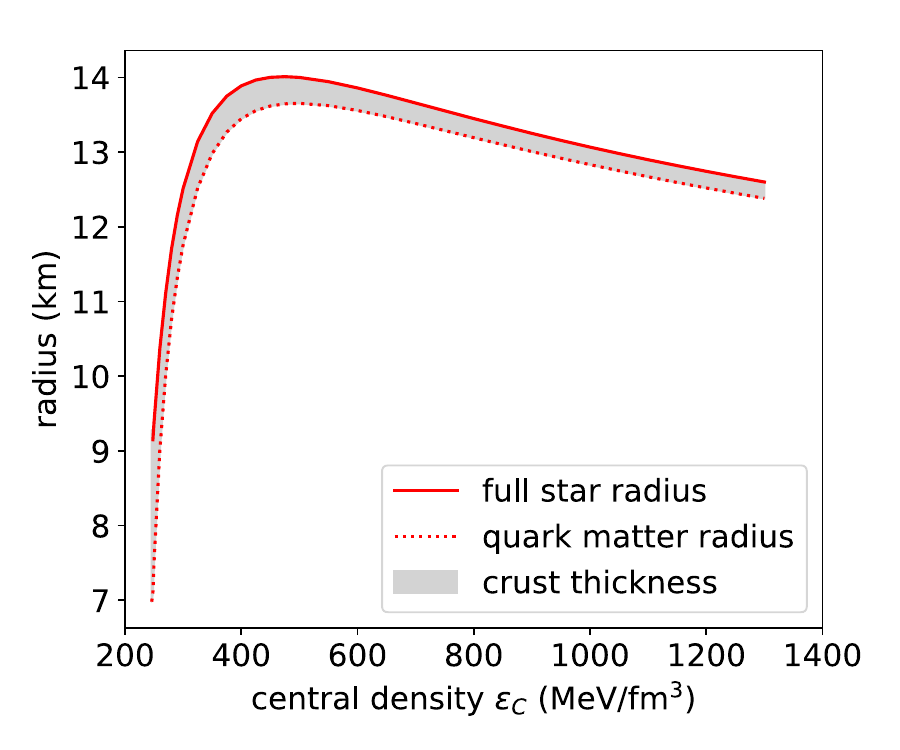}
   \caption{The thickness of the nuclear crust of quark stars, with quark matter modeled using the MIT Bag Model, featuring a bag constant of $B = 45$ MeV/fm$^3$.}
   \label{fig:crust_thick}
\end{figure}

The strange quark star sequences depicted in Fig.~\ref{fig:prns_quark} are calculated for bare quark stars, consisting entirely of strange quark matter from the core to the surface. As discussed by \citet{alcock1986, alcock1986b}, strange quark stars however do not necessarily have to be bare; they can be surrounded by nuclear crusts, with densities at the crust's base lower than neutron drip density. Examples of strange stars with nuclear crusts are presented in Fig.~\ref{fig:quark_crust}. The thickness of the nuclear crust of the compact members of the stars shown there is typically just a few hundred meters or less (see Fig.~\ref{fig:crust_thick}) and the mass is below $\sim 10^{-5} {\rm M}_\odot$ \citep{alcock1986}. However, rapid rotation can moderately increase these values, as demonstrated by \citet{Glendenning:1992crust}.
The value of the mass peaks for both the non-rotating and uniformly rotating sequences for quark matter are given in units of M$_\odot$ in Table~\ref{tab:rotation_quark} along with their corresponding radii and central energy density values. Uniform rotation at the Kepler limit for quark matter modeled within the MIT Bag Model increases the mass peak by 18\% when compared to the non-rotating counterpart, a larger increase than that of the DD2 model.

\begin{table}[]
\centering
\caption{Maximum gravitational mass with corresponding radius and central density of non-rotating and uniformly rotating quark stars for quark matter modeled within the MIT Bag Model with a bag constant $B = 45$~MeV/fm$^3$. }
\begin{tabular}{l c c c}
\toprule
  Rotation & Max. Mass [M$_\odot$] & Radius [km] & $\epsilon_c$ [MeV/fm$^{3}$]  \\ \colrule
  Non-rotating (TOV)   & 2.047 & 11.217 & 1100  \\
  Uniform ($\Omega_K$) & 2.413 & 13.250 & 900   \\ \botrule
\end{tabular}
\label{tab:rotation_quark}
\end{table}

\begin{figure}[htb]
\centering
   \includegraphics[width=8.5cm]{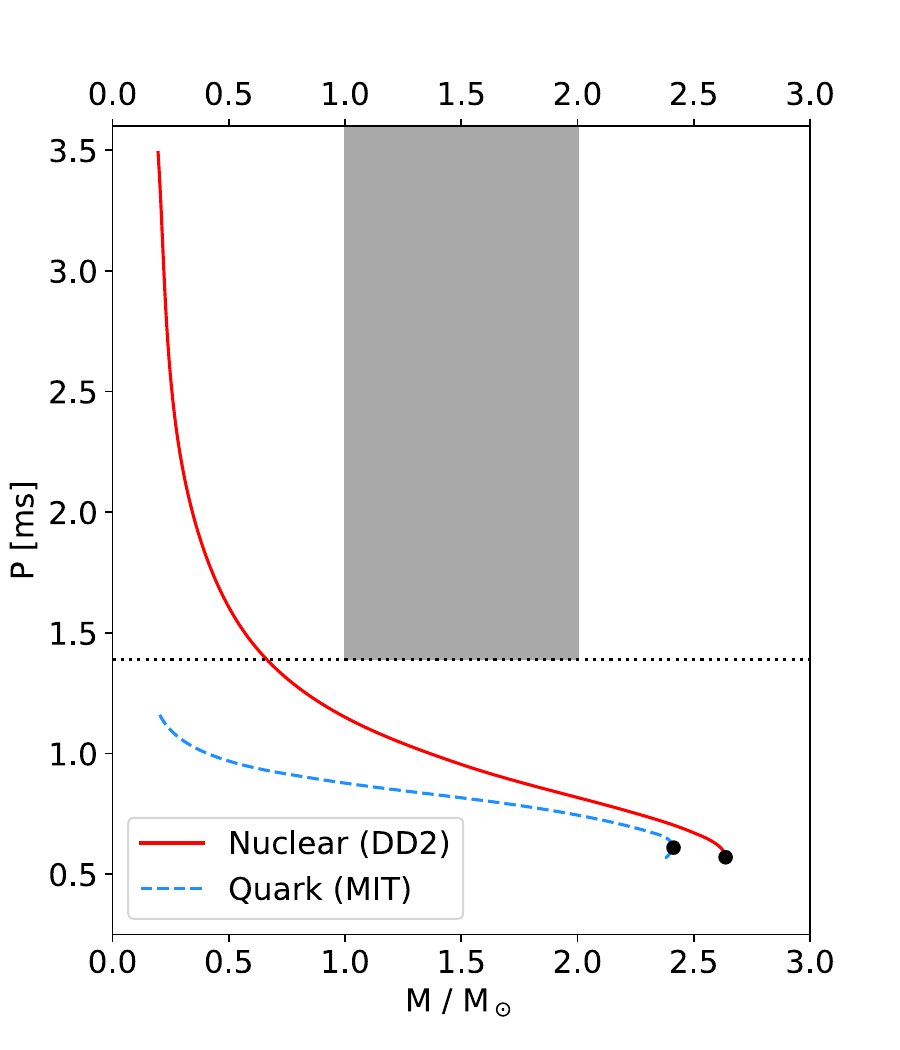}
   \caption{Rotational periods $P$ (in milliseconds) versus gravitational mass of both neutron stars and strange quark stars. Neutron stars are calculated using the DD2 nuclear EOS, while strange stars are modeled within the MIT Bag Model with the coupling constants $\alpha_s$ = 0, the superfluid gap $\Delta$ = 0 MeV, and a bag constant of $B = 45$ MeV/fm$^3$. The black dotted line represents the period of the fastest observed MSP ($P = 1.3889$ ms), and the grey-shaded region denotes the range encompassing all observed pulsars to date.}
   \label{fig:rot_period}
\end{figure}

The rotational period, $P$, of the uniformly rotating sequences from both the DD2 model and the MIT Bag Model can also be seen in Figure~\ref{fig:rot_period}. At lower mass values, $P$ is much higher for the nuclear EOS model DD2, but at their mass peaks, both the nuclear and quark EOS models see a similar value for $P$. In Figure~\ref{fig:rot_period}, the horizontal black dotted line indicates the rotational period of the fastest observed MSP, PSR J1748–2446ad, with a rotational period of 1.3889~ms.
The grey-shaded region schematically shows the period-mass range of all 
observed pulsars to date. The position of this region will shift downwards to shorter rotational periods if future pulsar discoveries unveil objects rotating below 1.3899~ms. Particularly, super-rapid rotation below approximately 1~ms, contingent on the mass, could suggest the presence of strange quark stars.

\subsection{Differential Rotation}\label{sec:dr}

While the majority of pulsars are believed to be rotating uniformly, some neutron stars form with an appreciable amount of differential rotation following extreme astrophysical events like core-collapse supernovae or binary neutron star mergers. 

A newly formed proto-neutron star may experience some degree of differential rotation following a core-collapse supernova, but the phenomena would be short-lived. Hydrodynamical forces in the form of Ekman pumping and turbulent mixing will damp out any differential rotation in the star within a few days \citep{Weber:1999book}. A more likely avenue for a star to experience differential rotation is following a binary neutron star merger event. When a neutron star binary coalesces, the two stars are most likely irrotational just prior to merging. The collision would result in a substantial velocity discontinuity at the surface of contact, giving rise to differentially rotating hypermassive or supramassive remnant stars. The remnant stars' fate depends largely on the dynamics and mass of the binary prior to merging. Many previous studies \citep{morrison2004effect, shapiro2000differential, baumgarte1999maximum, gondek2017new} have explored the implications of massive, differentially rotating neutron stars, but we extend this line of research to also include differentially rotating strange quark stars.

Differential rotation has physical implications that allow stars to remain stable in otherwise unstable configurations. Most notably, differentially rotating stars can support more mass than their uniformly rotating or non-rotating counterparts \citep{morrison2004effect}. In the case of massive remnant stars of binary merger events, differential rotation is one of the mechanisms that provide extra centrifugal support to stabilize the star well above the TOV and uniform rotation mass limits. 
    
The modeling of differential rotation begins with the same metric as uniform rotation. Once solved for, the metric functions are used to solve the equation of hydrostatic equilibrium for a barotropic fluid:
\begin{equation}
    h(P) - h_p = \frac{1}{2} [\gamma_p + \rho_p - \gamma - \rho - \text{ln}(1-v^2) + F(\Omega)]
    \label{hydro}
\end{equation}
where $h(P)$ is the enthalpy as a function of pressure, and $\gamma_p$ and $\rho_p$ are the values of the metric potentials at the pole. The quantity $v$ is defined as
\begin{equation}
    v = (\Omega - \omega) \, r \, \sin \, \theta e^{-\rho},
\end{equation}
and the last term, $F(\Omega)$, defines the rotational law for the matter. Following the form in \citep{komatsu1989rapidly,cook1992spin,cook1994rapidly},  the rotational law is set to
\begin{equation}
    F(\Omega) = A^2 (\Omega_c - \Omega),
    \label{rotlaw}
\end{equation}
where $\Omega_c$ is the central value for the angular velocity. In this equation, the parameter $A$ is a constant scaling factor of the degree of differential rotation and is used to determine the length scale over which the frequency changes \citep{morrison2004effect}. 


Simplifying Eq.~(\ref{hydro}) and including the rotation law gives 
\begin{equation}
    (\Omega_c - \Omega) = \frac{1}{A^2} \left[ \frac{(\Omega-\omega)\, s^2 (1-\mu^2) \,e^{-2\rho}}{(1-s)^2-(\Omega-\omega)^2\, s^2\, (1-\mu^2)\, e^{-2\rho}} \right],
\end{equation}\label{eq:ahat}
where the rotational frequency $\Omega$ can be isolated and solved numerically using a root-finding algorithm. The other non-defined parameters, $s$ and $\mu$, represent unitless radial and polar coordinates. Equation~(\ref{eq:ahat}) uses a modified version of the rotation parameter, labeled $\hat{A}$, which is scaled by the equatorial radius $r_e$: $\hat{A} \equiv A/r_e$. Uniform rotation is achieved in the limit $\hat{A}^{-1} \rightarrow 0$, and has an upper bound of 1. The gravitational and matter fields are solved in a self-consistent fashion; once a set of self-consistent solutions to these equations is obtained, physical quantities like mass, radius, angular momentum, and rotational kinetic energy are outputted.

\subsubsection{Results for Differentially Rotating Compact Objects}\label{sec:dr_ns}

In this section, M-R relations for differentially rotating neutron stars and quark stars are shown. The M-R relations are calculated for different levels of differential rotation, varying the rotation parameter $A$ in Eq.~(\ref{rotlaw}). Because the rotation parameter $A$ repeatedly appears as $\hat{A}^{-1}$ in the equations described in Section~\ref{sec:dr}, we follow the lead of previous work which parameterized sequences by values of $\hat{A}^{-1}$ = 0.3, 0.5, 0.7, and 1.0 \citep{morrison2004effect, cook1992spin, galeazzi2012differentially}. It's important to note that using the numerical scheme described in Section~\ref{sec:dr}, the most extreme case of differential rotation ($\hat{A}^{-1}$ = 1.0) may fail to find solutions at high central densities depending on the underlying EOS. 
\begin{figure}[tb]
\centering
   \includegraphics[width=8.5cm]{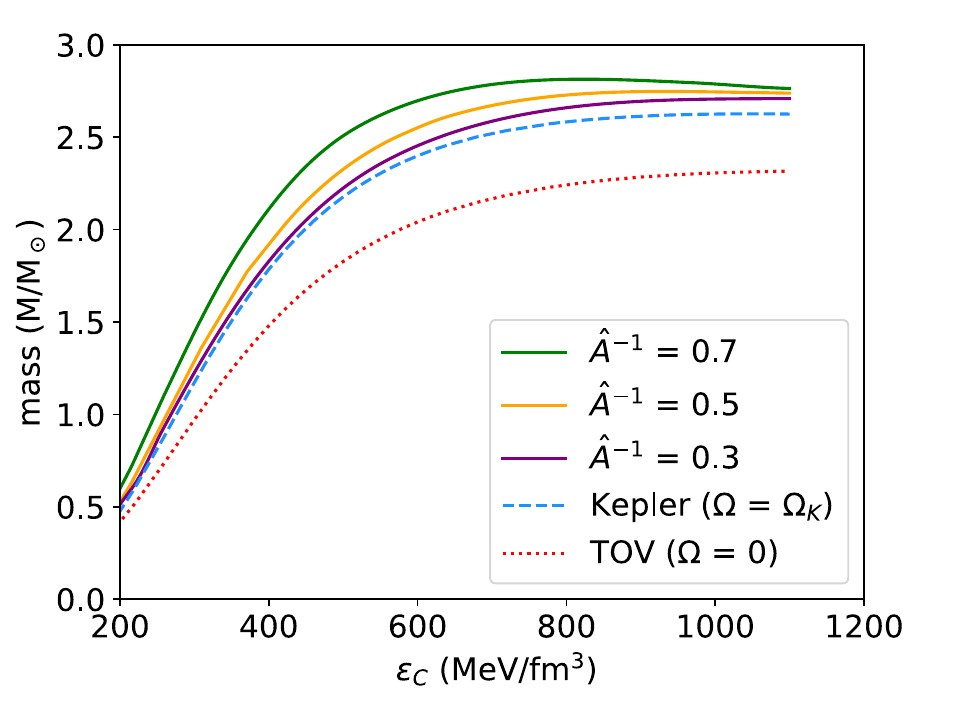}
   \caption{Mass as a function of central density $\epsilon_c$ for the DD2 parameter set, comparing three degrees of differential rotation characterized by the rotation parameter $\hat{A}^{-1}$ to the uniformly rotating (Kepler) and non-rotating (TOV) cases.}
   \label{fig:dr_dd2}
\end{figure}
In Fig.~\ref{fig:dr_dd2}, the relationship between mass and central density $\epsilon_c$ is illustrated for the DD2 parameter set. This comparison involves three levels of differential rotation characterized by the rotation parameter $\hat{A}^{-1}$, contrasting them with the uniformly rotating (Kepler) and non-rotating (TOV) scenarios discussed in Section~\ref{subsec:uniform}. The figure underscores the influence of varying degrees of differential rotation on the mass-central density relationship. 
\begin{table}[tb]
\centering
\caption{Maximum gravitational mass with corresponding radius and central density of differentially rotating neutron stars, parametrized by three degrees $\hat{A}^{-1}$ of differential rotation, modeled with the DD2 parameter set.}
\begin{tabular}{c c c c }
\toprule
  $\hat{A}^{-1}$ & Mass [M$_\odot$] & Radius [km] & $\epsilon_c$ [MeV/fm$^{3}$]  \\ \colrule
  0.3  & 2.66 & 13.05 & 1100    \\
  0.5  & 2.73 & 13.35 & 920   \\ 
  0.7  & 2.75 & 13.59 & 760  \\ \botrule
\end{tabular}
\label{tab:rotation_dr_hadronic}
\end{table}
Table~\ref{tab:rotation_dr_hadronic} provides detailed information on the maximum gravitational mass, corresponding radius, and central density for differentially rotating neutron stars. These models are parametrized by three levels of the rotation parameter $\hat{A}^{-1}$: 0.3, 0.5, and 0.7, as modeled with the DD2 parameter set. The results demonstrate how the introduction of differential rotation affects the stellar properties, with decreasing values of $\hat{A}^{-1}$ gradually converging toward the uniform rotation scenario. This analysis highlights the nuanced impact of differential rotation on the stellar structure and provides a comprehensive understanding of the mass-central density relationship for neutron stars with varying degrees of rotation.

We next turn to investigating the impact of differential rotation on quark stars. Following the description in Section~\ref{quark_eos_sec}, six equation of state models are constructed: three models of quark matter with corrections due to the strong coupling constant, and three models of quark matter in the CFL phase. The three models with corrections due to the strong coupling constant, $\alpha_s$, vary in the value of $\alpha_s$ (0.0, 0.2, and 0.35). The other three CFL models vary the parameter $\Delta$, setting equal to 0, 50, or 100. For all six models. the bag constant $B$ is set to 45~MeV/fm$^3$.

\begin{figure}[tb]
\begin{minipage}{\linewidth}
  \centering
  \includegraphics[width=6.5cm]{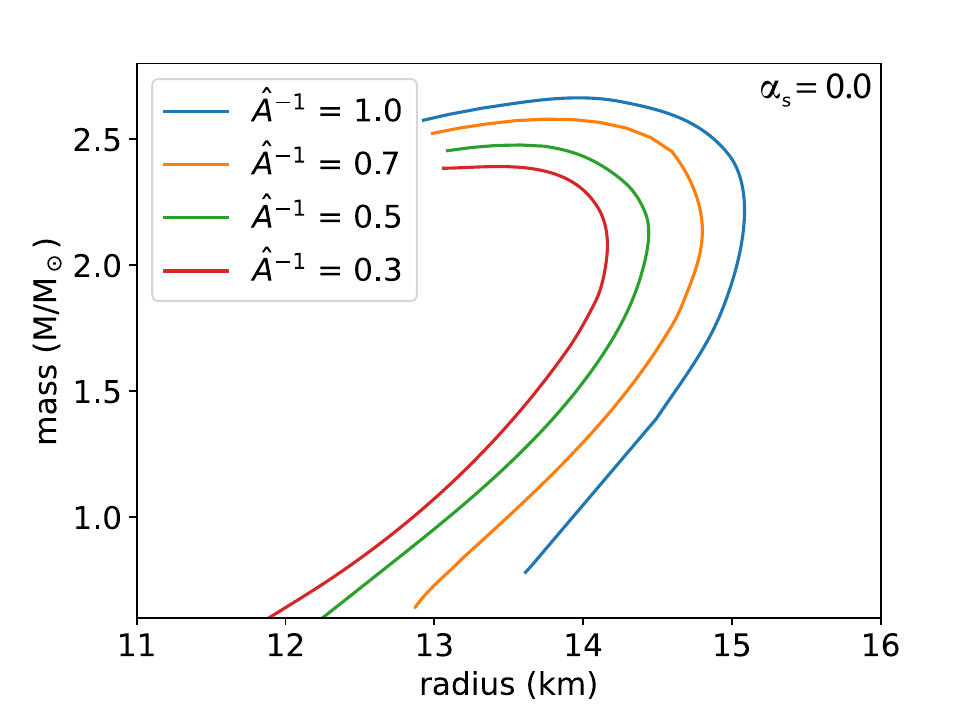}
    \label{fig:alpha00}
\end{minipage}
\vspace{1em} 
\begin{minipage}[b]{0.45\linewidth}
  \centering
  \includegraphics[width=6.5cm]{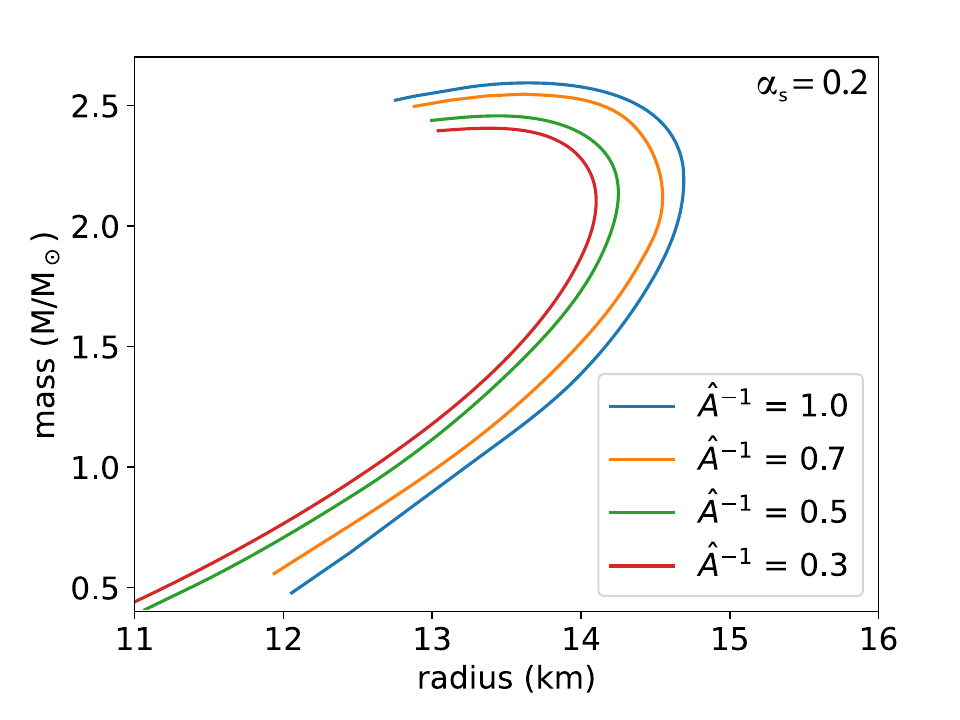}
  \label{fig:alpha02}
\end{minipage}
\hfill
\begin{minipage}[b]{0.45\linewidth}
  \centering
  \includegraphics[width=6.5cm]{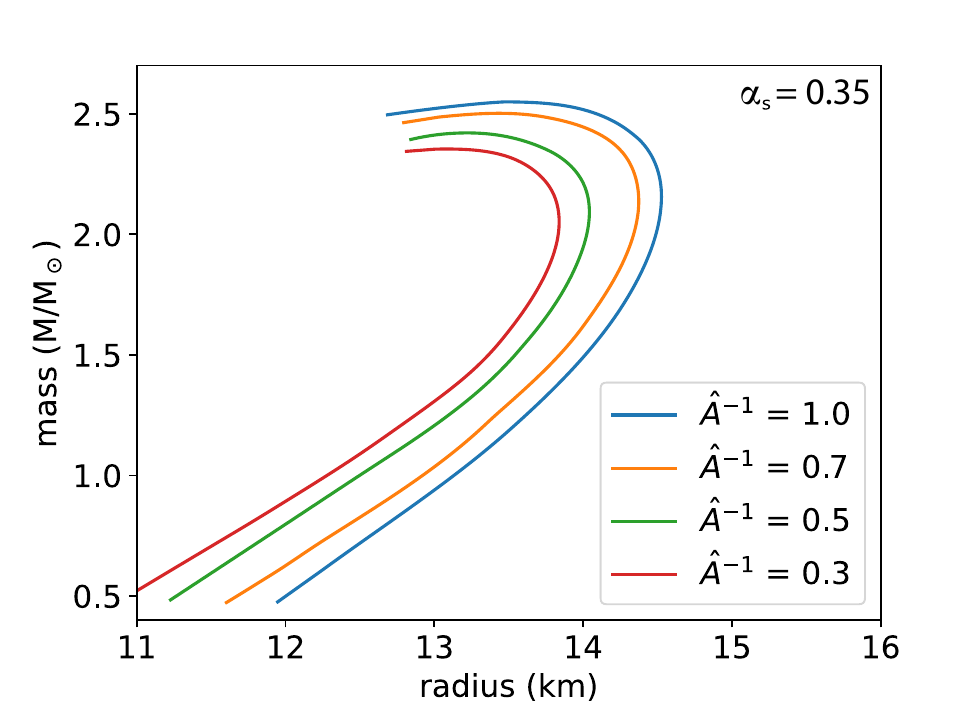}
  \label{fig:alpha035}
\end{minipage}
\caption{Mass-radius relations of quark stars for the MIT Bag Model EOS with three values of the strong coupling constant $\alpha_s$ (0.0 on top, 0.2 on the bottom left, and 0.35 on the bottom right). Each figure depicts four degrees of differential rotation characterized by the scaled rotation parameter $\hat{A}^{-1}$.}
\label{dr_mit}
\end{figure}

\begin{figure}[tb]
\begin{minipage}{\linewidth}
  \centering
  \includegraphics[width=6.5cm]{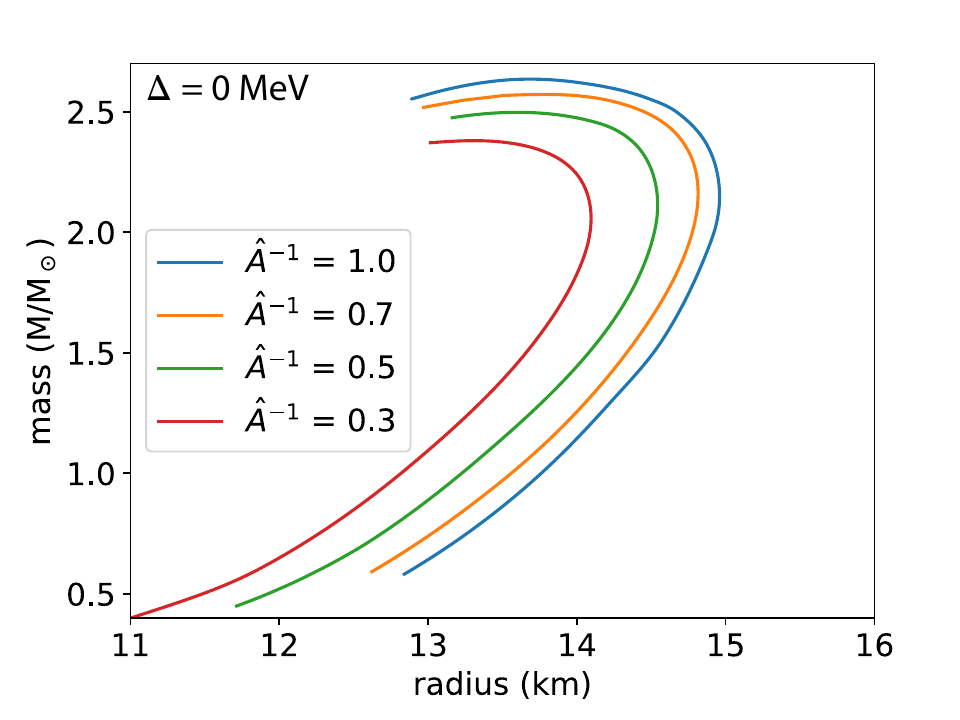}
    \label{fig:delta0}
\end{minipage}
\vspace{1em} 
\begin{minipage}[b]{0.45\linewidth}
  \centering
  \includegraphics[width=6.5cm]{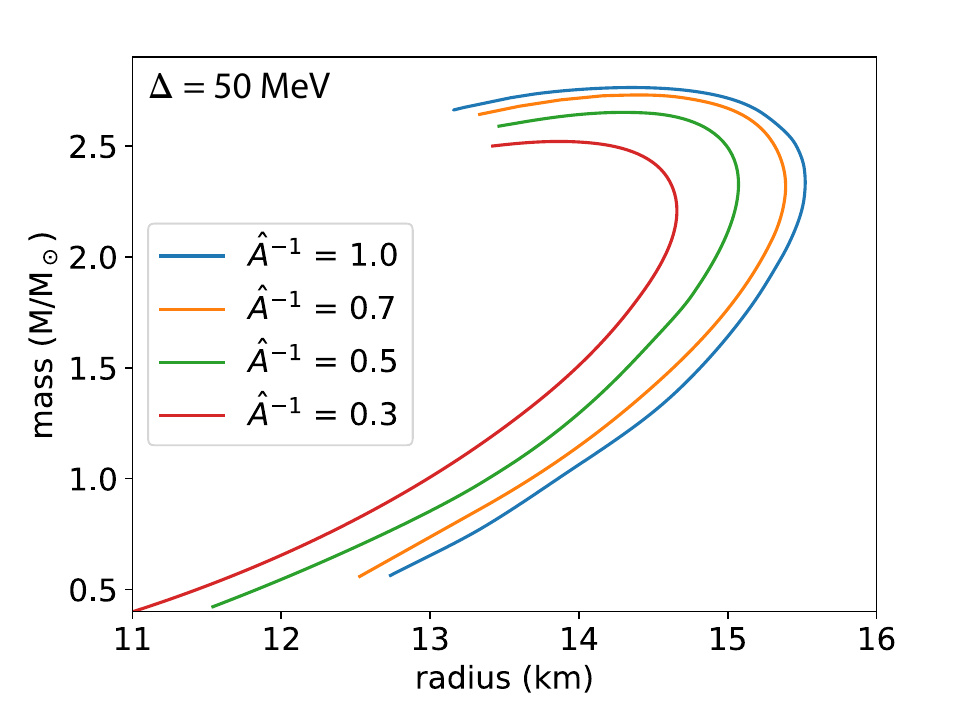}
  \label{fig:delta50}
\end{minipage}
\hfill
\begin{minipage}[b]{0.45\linewidth}
  \centering
  \includegraphics[width=6.5cm]{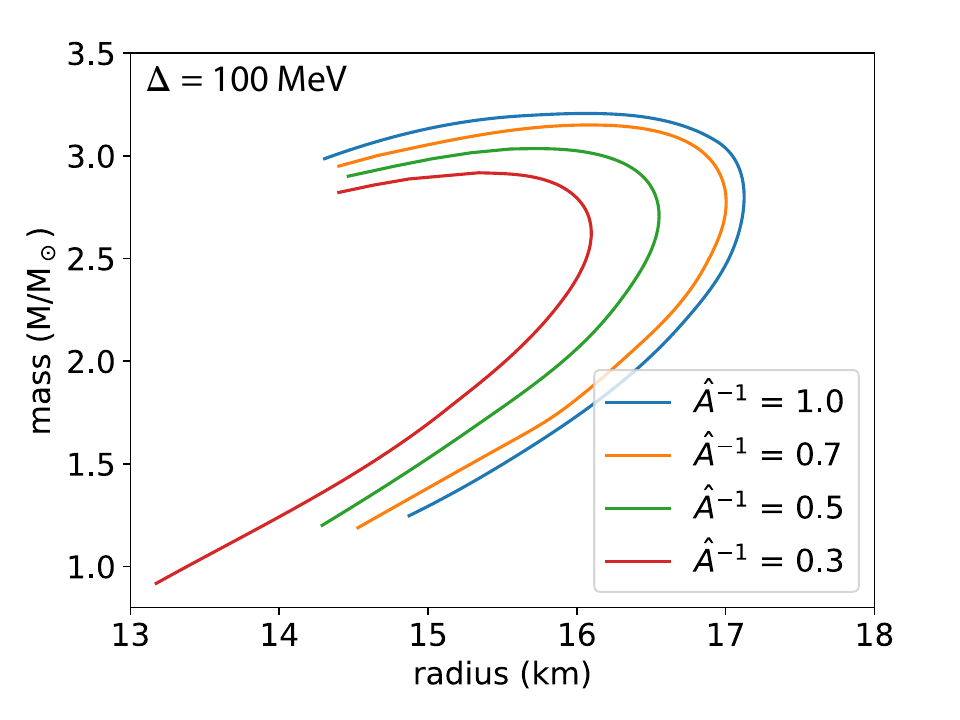}
  \label{fig:delta100}
\end{minipage}
\caption{Mass-radius relations of quark stars computed for the MIT Bag Model EOS with three values of $\Delta$ in the CFL phase (0 MeV on top, 50 MeV on the bottom left, and 100 MeV on the bottom right). Each figure depicts four degrees of differential rotation characterized by the scaled rotation parameter $\hat{A}^{-1}$.}
\label{dr_cfl}
\end{figure}

M-R relations for four different values of the rotation parameter $\hat{A}^{-1}$ (0.3, 0.5, 0.7, and 1.0) are constructed for each EOS model. The M-R relations varying $\alpha_s$ are shown visually in Figure~\ref{dr_mit}, and the M-R relations varying $\Delta$ are shown visually in  Figure~\ref{dr_cfl}. The maximum gravitational mass from each curve for the four values of $\hat{A}^{-1}$ are given in Table~\ref{dr_max_mass}. The general trend for maximum mass expected when varying the value of $\hat{A}^{-1}$ is to see higher mass values as $A^{-1} \rightarrow 1.0$, which is reflected in Table~\ref{dr_max_mass}. Additionally, stellar models with higher degrees of differential rotation struggle to find physical results at lower masses, as reflected in Figures~\ref{dr_mit} and \ref{dr_cfl} for both sets of quark matter EOS models. 

The mass-radius relations in Figures~\ref{dr_mit} and \ref{dr_cfl} are constructed for a range of constant central densities $\epsilon_c$. Two 
\begin{table}[tb]
\centering
\caption{Maximum gravitational mass of quark stars (in units of M$_\odot$) for four degrees of differential rotation characterized by the rotation parameter $\hat{A}^{-1}$. The first three rows vary the strong coupling constant $\alpha_s$, and the last three vary the superfluid gap $\Delta$ in the CFL phase.}
\begin{tabular}{l c c c c }
\toprule
  Parametrization & $\hat{A}^{-1}$ = 0.3 & $\hat{A}^{-1}$ = 0.5 & $\hat{A}^{-1}$  = 0.7 & $\hat{A}^{-1}$ = 1.0 \\ \colrule
  $\alpha_s$ = 0.0   & 2.384 & 2.476 & 2.579 & 2.664 \\
  $\alpha_s$ = 0.2   & 2.391 & 2.443 & 2.546 & 2.597 \\
  $\alpha_s$ = 0.35  & 2.353 & 2.411 & 2.506 & 2.549 \\ \colrule
  $\Delta$ = 0 MeV   & 2.379 & 2.492 & 2.576 & 2.651 \\
  $\Delta$ = 50 MeV  & 2.525 & 2.647 & 2.732 & 2.767 \\
  $\Delta$ = 100 MeV & 2.917 & 3.037 & 3.162 & 3.193 \\ 
\botrule
\end{tabular}
\label{dr_max_mass}
\end{table}
additional parameters are specified for each curve: the ratio of polar to equatorial ratio for the stellar models r$_{\text{ratio}}$ and the given rotation parameter $\hat{A}^{-1}$. As the magnitude of differential rotation increases in a star, signified by $\hat{A}^{-1} \rightarrow 1$, the star grows more deformed, expanding at the equator, and thus decreases the r$_{\text{ratio}}$. For extremely high levels of differential rotation in massive stars, the star may even form the shape of a torus as r$_{\text{ratio}} \rightarrow 0$. The numerical scheme described in Section~\ref{sec:dr} uses 
the central density as the maximum density within the star which limits the lower threshold for the parameter r$_{\text{ratio}}$, so only quasi-toroidal configurations are possible for extreme differential rotation. The reader can find more information on modifying the specified numerical scheme to include fully toroidal configurations in \citet{kaplan2014influence} and  \citet{morrison2004effect}.

\section{Summary and Concluding Remarks}

This chapter covers various topics related to neutron stars, addressing historical challenges in measuring their radius and the consequent uncertainties in size determination. Methods for radius estimation involve spectroscopic and timing techniques, with considerations for factors like magnetic fields and rapid rotation. Pulsar glitches or sudden spin jumps in neutron stars, provide insights into dense matter's EOS, revealing details about angular momentum transfer and interior dynamics. Tidal deformability, crucial in gravitational-wave astronomy, is explored in the context of binary neutron star mergers, with emphasis on the GW170817 event's contributions to mass, radius, and deformability estimates. The discussion extends to fast pulsars, categorizing them based on rotation, magnetic fields, and the role they play in astrophysical discoveries. The EOS of neutron stars is detailed, encompassing lower-density crust models and high-density core approaches. The chapter concludes by delving into quark stars, strange dwarfs, and the transformative possibilities of understanding these exotic objects through future observations and gravitational wave detectors like LISA. Additionally, properties of rotating neutron stars are thoroughly examined, covering uniform and differential rotation scenarios, with insights into the impact of rotation on mass and radius, particularly after binary neutron star mergers. The comparison of maximum gravitational masses offers valuable information about the stability and deformations of rotating compact objects.

Key insights derived from the examination of fast pulsars, neutron stars, and astrophysical objects featuring strange quark matter include the following:
\begin{enumerate}
    \item Historical challenges related to direct measurements of neutron star radius have resulted in uncertainties, necessitating the reliance on spectroscopic and timing techniques. Spectroscopic methods, particularly in low-mass X-ray binary systems, involve making assumptions about emitted spectra to infer mass-radius relations. However, challenges arise in cases of strong magnetic fields or rapid rotation.
    \item Pulsar glitches offer a unique perspective into the equation of state of dense neutron star matter, as the proposed models involve angular momentum transfer between the solid crust and the extremely dense interior, specifically the neutron superfluid.
    \item Tidal deformability, crucial in gravitational-wave astronomy, is elaborated in the context of binary neutron star mergers. Theoretical calculations, along with the groundbreaking GW170817 event, contribute to estimates of masses, radii, and dimensionless tidal deformability.
    \item Fast pulsars have played a pivotal role in testing theories of relativistic astrophysics. With over 3,000 observed pulsars to date, their numbers continue to grow rapidly with advancements in telescopes. Measuring pulsar rotation provides essential constraints on their bulk properties and the nature of the matter in their central cores. Superfast rotation, ideally with rotational periods below 1 millisecond, is expected to provide an additional observational window into the inner workings of neutron stars.
    \item The equation of state (EOS), governing the pressure-density relationship within neutron stars, is crucial for understanding their internal structure. Models for the lower-density crust and high-density core are briefly discussed, emphasizing their role in simulations and observational analyses.
    \item The intriguing concept of strange quark stars is explored. If they exist, such stars could be composed of quark matter in the color-flavor-locked phase, among other condensation patterns. The properties of strange dwarfs, characterized by conventional nuclear crusts and strange quark matter cores, are reviewed. Recently, potential implications for gravitational wave astronomy have been highlighted by \citet{Perot.universe:2023}. 
    
    The discovery of stellar objects composed of absolutely stable strange quark matter remains an intriguing and challenging prospect in astrophysics. Detecting such objects presents a significant challenge due to the numerous observational signatures that overlap with those of ordinary neutron stars and white dwarfs. Distinguishing features may arise from ultra-rapid rotation (well below 1~ms), small radii (less than 10~km), and photon emission from the surface of a bare strange star. Proposed candidates, like the compact star RX J1856.5-3754, PSR J1614-2230, PSR J0348+0432, 4U 1820-30, and HESS J1731-347 have been considered, but definitive evidence is yet to be established. Advanced observational techniques and future space-based observatories, coupled with refined theoretical models, offer hope in the ongoing quest to unravel the mysteries of strange quark matter and identify potential stellar objects made of such matter.
    \item Theoretical frameworks for calculating properties of rotating neutron stars are presented, covering uniform and differentially rotating scenarios. The impact of rotation on mass and radius is discussed using the DD2 model for the nuclear equation of state. Insights into stability and deformations are gained through comparisons of maximum gravitational masses.
\end{enumerate}

\noindent
\section*{Acknowledgments}
  This research was supported by the National Science Foundation (USA)
  under Grant No.\ PHY-2012152. MGO, IFR-S, and MC thank CONICET,  
  UNLP and MinCyT (Argentina) for financial support under grants
  PIP-0169, 11/G187 and PICT 2019-3662. MC thanks Dr. Mauro Mariani and Marcos Celi for their assistance in creating the 3D figures for this chapter.


\printindex

\end{document}